\documentclass[twocolumn,amsmath,amssymb,aps,prc,superscriptaddress]{revtex4-2}

\usepackage{xcolor}
\usepackage[utf8]{inputenc}
\usepackage[T1]{fontenc}
\usepackage{subfig}
\usepackage{graphicx}
\DeclareGraphicsExtensions{.pdf,.jpeg,.png}
\graphicspath{./}
\usepackage{enumerate}
\usepackage{slashed}
\usepackage{textcomp}
\usepackage{mathpazo}
\usepackage{natbib}
\usepackage{array}
\usepackage[normalem]{ulem}
\usepackage{multirow}
\usepackage{bm}	 
\usepackage[colorlinks=true,allcolors=blue]{hyperref} 

\begin{document}

\title{Bottomonium suppression in PbPb collision at energies available at the CERN large hadron collider}

\author{Nikhil Hatwar}
\email{nikhil.hatwar@gmail.com}
\affiliation{Department of Physics, Birla Institute of Technology and Science, Pilani, Rajasthan, India}
\author{Captain R. Singh}
\affiliation{Department of Physics, Indian Institute of Technology, Indore, Madhya Pradesh, India}
\author{S. Ganesh}
\noaffiliation{}
\author{M. Mishra}
\email{madhukar@pilani.bits-pilani.ac.in}
\affiliation{Department of Physics, Birla Institute of Technology and Science, Pilani, Rajasthan, India}

\date{\today}

\begin{abstract}

High energy collisions are the laboratories within our reach to study the strongly interacting matter under extreme temperatures. In the present study, we use a quarkonia suppression scheme to explain the bottomonium production at the two LHC energies. We employ ECHO-QGP to model the ($3+1$)-dimensional relativistic viscous hydrodynamic evolution of the medium. Bottomonia produced in the early stage dissociates due to color screening, gluonic dissociation, and collisional damping in addition to the shadowing as an initial state effect. In the color screening mechanism, the temperature from hydrodynamics is used to find the screening radii at each centrality and rapidity. Shadowing effect utilizes the parton distribution functions obtained from {\it CT14} global analysis and shadowing factors from {\it EPPS16}. A Lattice QCD based equation of state from Wuppertal-Budapest collaboration has been used. The experimental values of poins($\pi^+$) spectra were used to constrain the initial conditions of the dynamics. The bottomonium suppression is determined as a function of centrality, transverse momentum and rapidity for $\Upsilon(1S)$ and $\Upsilon(2S)$ at the LHC energies of $2.76$ TeV and $5.02$ TeV. We find a fairly good agreement between our theoretically calculated survival probability and the measured nuclear modification factor($R_{AA}$) at the two energies.
\end{abstract}
\maketitle

\section{Introduction}

We have made considerable progress in our understanding of the strong interaction since the advent of heavy-ion collision experiments especially at Relativistic Heavy-ion Collider(RHIC) and at Large Hadron Collider (LHC). Quantum Chromodynamics (QCD), the theory of strong interaction predicts that a thermalized medium of Quark-Gluon Plasma (QGP) should form as a transient stage of heavy-ion collisions when the temperature and/or matter density of the fireball exceeds a certain threshold value.
The largest portion of the speculative phase diagram of the QCD is occupied by this QGP~\cite{jacak2012exploration,guenther2021overview}. 
It is pertinent for us to investigate the nature of this medium in order to make advancement in our understanding of QCD and its limits.
And with consensus that QGP does exist in ultra-relativistic energy nucleus-nucleus collisions, the focus is now shifting towards characterizing this medium with the help of the probes which acted as its signatures. A good starting point is to use the wide range of available experimental data from different systems of collisions and energies to constrain the input parameters of the dynamics at work.
 
 So far, the experimental findings indicate the formation of an inviscid medium~\cite{adams2005experimental,krzewicki2011elliptic}. 
A theoretical lower limit for the shear viscosity to entropy density ratio($\eta/s$) of strongly interacting matter has been know for a while~\cite{kovtun2005viscosity}. 
But the estimation of such low $\eta/s$ for quark matter in heavy ion collision has become possible only recently using lattice QCD~\cite{ratti2018lattice} and \textit{bayesian parameter estimation methods}~\cite{bernhard2019bayesian}.
Hydrodynamics has been widely applied to simulated heavy ion collisions of various systems and energies.
Phenomenological models based on hydrodynamics better explain the observables from high energy collisions with thermalization started early, $\tau$<1fm/c ~\cite{heinz2004thermalization,martinez2008dilepton}. 
What makes this problem of characterizing QGP so difficult is the sheer complexity of the system. In high energy  collisions, the only variables in our control are particle species and the energy of collision. All the rest has to be inferred indirectly from the observables.
Predicting the transport and thermodynamic properties of a medium formed in the collision by analysing patterns in the produced particle yields is a scrupulous task. Hence, modelling of such a complex system can only be dealt phenomenologically~\cite{dubla2018towards}.
Right after the collision, we get a fireball where the medium consisting of quarks and gluons expands against the surrounding vacuum and cools down rapidly.
We assume that this system thermalizes quickly after the collision and we mark the time required for the system to thermalize as QGP formation time. 
Hydrodynamics is switched off when the system temperature falls below the pseudo critical temperature($T_c$) of the QGP. The recently agreed upon value of $T_c$ for QGP computed by Lattice QCD collaborations is $156$ MeV~\cite{bazavov2019chiral,steinbrecher2019qcd,d2019high}. 
One way of testing a given model is by calculating a physical quantity which could act as a theoretical counterpart of an observable measured by detectors. 
If QGP does exist as a transient stage of heavy-ion collision, then we should be able to notice an agreement between these two quantities~\cite{bass1999signatures}. 
Among many of such signatures of QGP, one is quarkonia suppression, on which we are concentrating in the current work. 
Quarkonia are mesonic bound states of heavy quark and heavy anti quark, which are produced in the early stage of collision. 
They could dissociate due to various types of interactions with the partons in the medium and would be detected comparatively lesser in number than in collision systems where we do not expect QGP, like at low energies and in p-p collisions~\cite{matsui1986}. 
In order to quantify this suppressed production of quarkonium, experimentalists measure a physical quantity called nuclear modification factor($R_{AA}$). 
It is the quarkonia yield in heavy-ion collision divided by the yield of the quarkonium in p-p collision scaled by $N_{coll}$. Its value less than one, greater than one and equal to one indicates suppression, enhancement and no medium effect, respectively. 
Surprisingly, there are cases which contribute to suppression in collision systems where we do not expect a thermalized medium.
These non-QGP effects arise due to situations before collisions (initial state effects where the system is said to be cold), even though small needs to be modelled into the total suppression scheme~\cite{aaij2014CMN}.
These non-QGP effects are called cold nuclear matter (CNM) effects. 
The $R_{AA}$ is measured over a wide range of collision energies as a function of centrality, $N_{part}$, transverse momentum, $p_T$ and rapidity, $y$. 
Various phenomenological models have tried to explain the measured values of the suppression consistently over a wide range of beam center of mass energies and collision systems~\cite{schukraft2017qm2017}. Few of them attempted explaining centrality and $p_T$ dependencies of ~\cite{vitev2013,Rapp2017}.
And even fewer predict all the three dependencies of the suppression over a wide range of available center of mass energies~\cite{strickland2018}. 
In our earlier work published in Eur. Phys. J C {\bf 79}, 147 (2019), we had explained the $p_T$ and $N_{part}$ dependence of $R_{AA}$ over a range of LHC energies. It was based on the suppression due to colour screening, gluonic dissociation and collisional damping under ($1+1$)-dimensional Bjorken's expansion of the thermalized medium. The net quarkonium yield was determined using a rate equation which combines suppression and recombination due to correlated quark anti quark pairs.

In the current study, we start with the initially produced bottomonia yields which evolves in ($3+1$)-dimensional hydrodynamic medium. This initial yield is influenced by a CNM effect used here called "Shadowing", which has been updated for the newly available parton distribution functions and shadowing factors. 
The bottomonium bound state while drifting through QGP could dissociate due to gluonic dissociation, collisional damping and color screening which are adapted to the ($3+1$)-dimensional hydrodynamic expansion. Color screening has been streamlined by eliminating the need of assuming a pressure profile of collision. A lattice QCD based equation of state(EOS) from  Wuppertal-Budapest Collaboration has been utilized. The input parameters for the hydrodynamics are constrained using the transverse momenta and rapidity spectra for poins from the ALICE experiment.   And lastly after considering the possibility of recombination of correlated bottom quark and anti-quark pair, we find the final number of bottomonia for ground state and excited states~\cite{digal2001quarkonium}.       
We then find a quantity called survival probability($S_p$) which is theoretically equivalent to the experimentally measured $R_{AA}$. We determine this as a function of transverse momentum, centrality and rapidity at $2.76$ TeV and $5.02$ TeV energies and then compare with the corresponding $R_{AA}$ values for $\Upsilon(1S)$ and $\Upsilon(2S)$ states. On comparison, we find a reasonably well agreement between $S_p$ and $R_{AA}$ at two LHC energies.\\

The arrangement of the topics in the current paper is as follows. Sec. I describes the general introduction of the proposed work. In Sec. II, we briefly describe the ($3+1$)-dimensional hydrodynamics used to model the bulk of medium using ECHO-QGP. It is followed by the quarkonia suppression formalism and the various effects incorporated into the scheme. Sec. III presents the results and discussions on the suppression of two bottomonium states, (1S) and (2S) versus $N_{part}$, $p_T$, and rapidity $y$ at the two LHC energies. Finally in Sec. IV, we summarize our results obtained and conclude the work.

\section{Formalism}
Here we describe the formalism in brief. The suppression formalism which has been developed in our previous work~\cite{singh2019}, has been adapted for the ($3+1$)-dimensional viscous hydrodynamics. More details about individual medium effects used can be found in~\cite{bottomganesh2013,qpmcolorscreening2013,MishraCS2007}. 

\subsection{(3+1)-dimensional Hydrodynamical Expansion of the Medium}

Hydrodynamics has been quite successful in explaining bulk observables from ultra-relativistic heavy-ion collisions for a wide range of system and energies~\cite{krouppa2019bottomonium,mcdonald2017hydrodynamic,alqahtani20173,habich2015particle}.
The agreement of hydrodynamical predictions with experimental results has been taken as an indirect evidence for the correctness of dynamics.   
The Bjorken's hydrodynamics assumes that the fireball expands only along the longitudinal direction and is restricted to the plateau region of rapidity spectra ($dN/dy$ vs $y$). This leads to the key variables of dynamics, e.g. temperature, pressure, energy density, entropy density to become an explicit function of proper time~\cite{bjorken1983highly}. Thus, although models based on Bjorken's evolution are adequate to estimate the observables at midrapidity, yet they are ineffective in providing the complete spacetime evolution of the system. In order to simulate the true dynamics of a collision that holds up expansion along the transverse directions and larger rapidities, one has to switch to the complete $(3+1)$-dimensional hydrodynamics. ECHO-QGP is a FORTRAN based code to find the solutions of the conservation equations, $d_{\mu}N^{\mu}=0$ and $d_{\mu}T^{\mu \nu}=0$, where, $d_\mu$ is the covariant derivative, $N^{\mu}$ is the four current and $T^{\mu \nu}$ is the energy momentum tensor. To solve these equations numerically in (3+1)-dimensions with relativistic speeds and viscous conditions, Israel-Stewart's second order formalism has been used in ECHO-QGP~\cite{del2013relativistic}.
A Cooper-Frye prescription handles the freeze-out stage, where the produced particles are assigned their momenta at the constant temperature hypersurface. We vary the input parameters of the ECHO-QGP so that the particle momentum spectra calculated here matches with the measured spectra from experiments as explained below.
We mark the end of the QGP phase at the value of proper time when the maximum temperature of the system drops below pseudo critical temperature, $T_c$.

\begin{figure}
\includegraphics[scale=0.27]{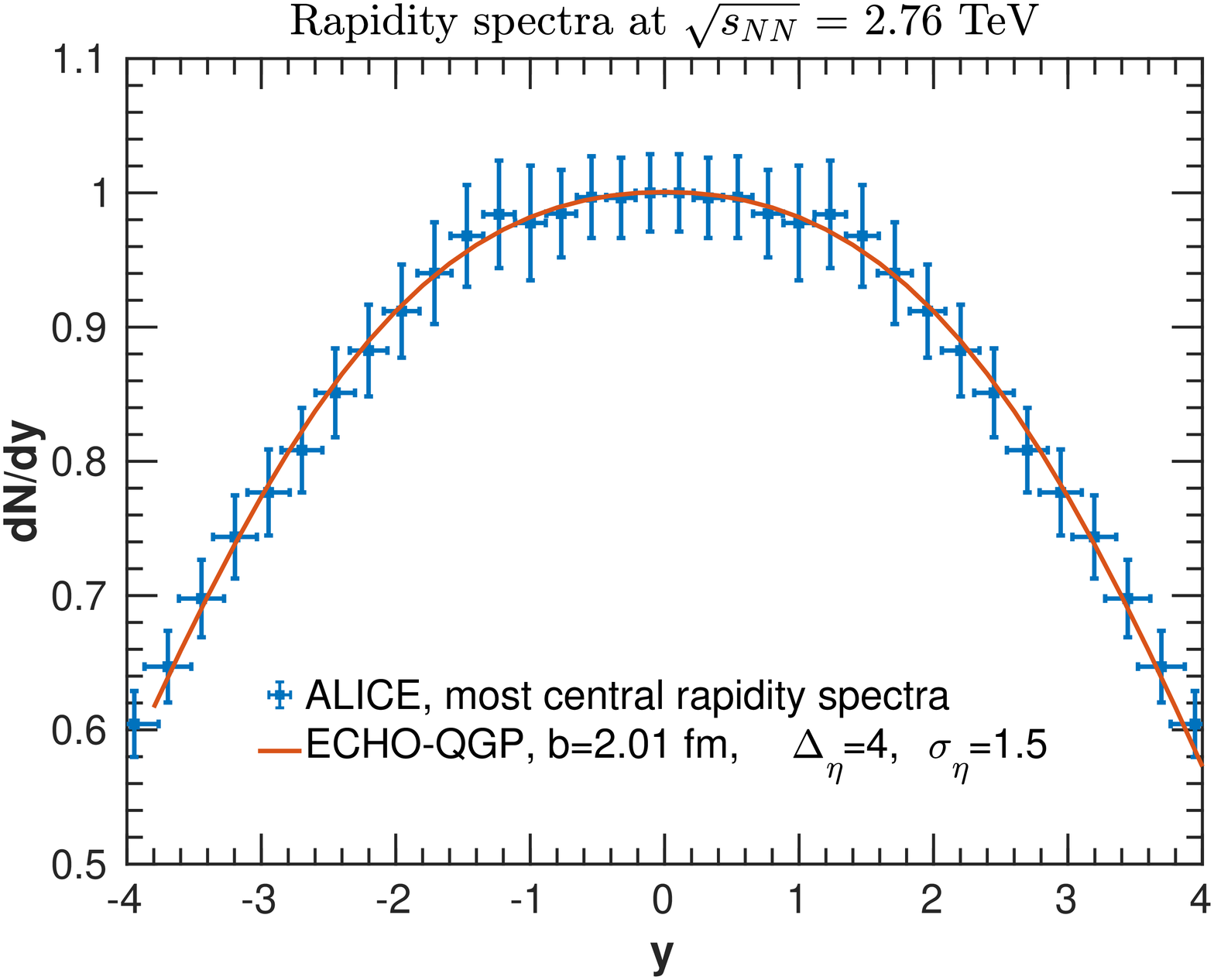}\\
\includegraphics[scale=0.27]{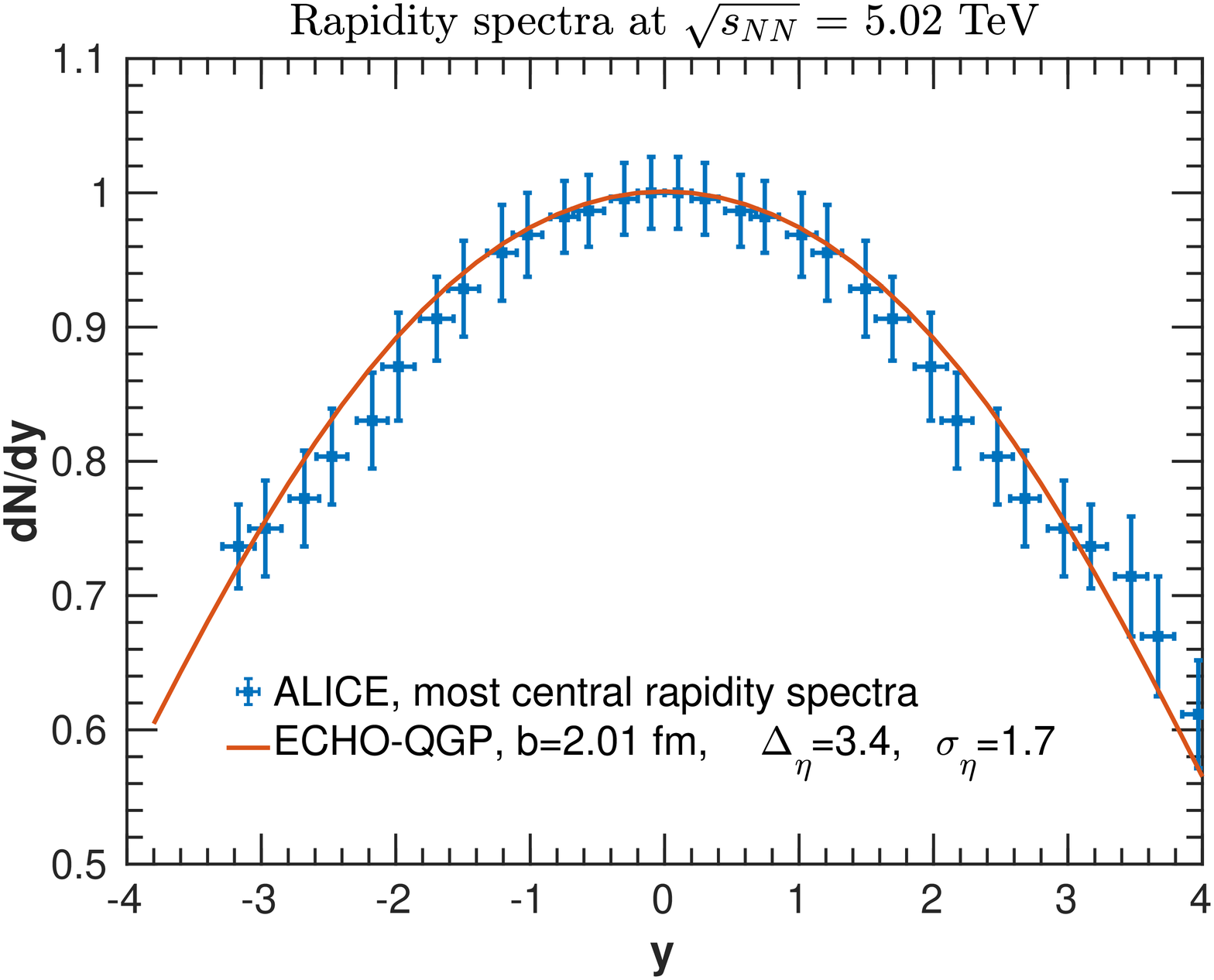}\\
\caption{Pion($\pi^+$) rapidity spectra for the two mentioned LHC energies from ALICE, normalized to their respective maximas for the most central collision(with impact parameter, \textit{b}) compared with the those obtained from ECHO-QGP.}
\label{fig:rap_spectra}
\end{figure}

\begin{figure}
\begin{center}
\includegraphics[scale=0.32]{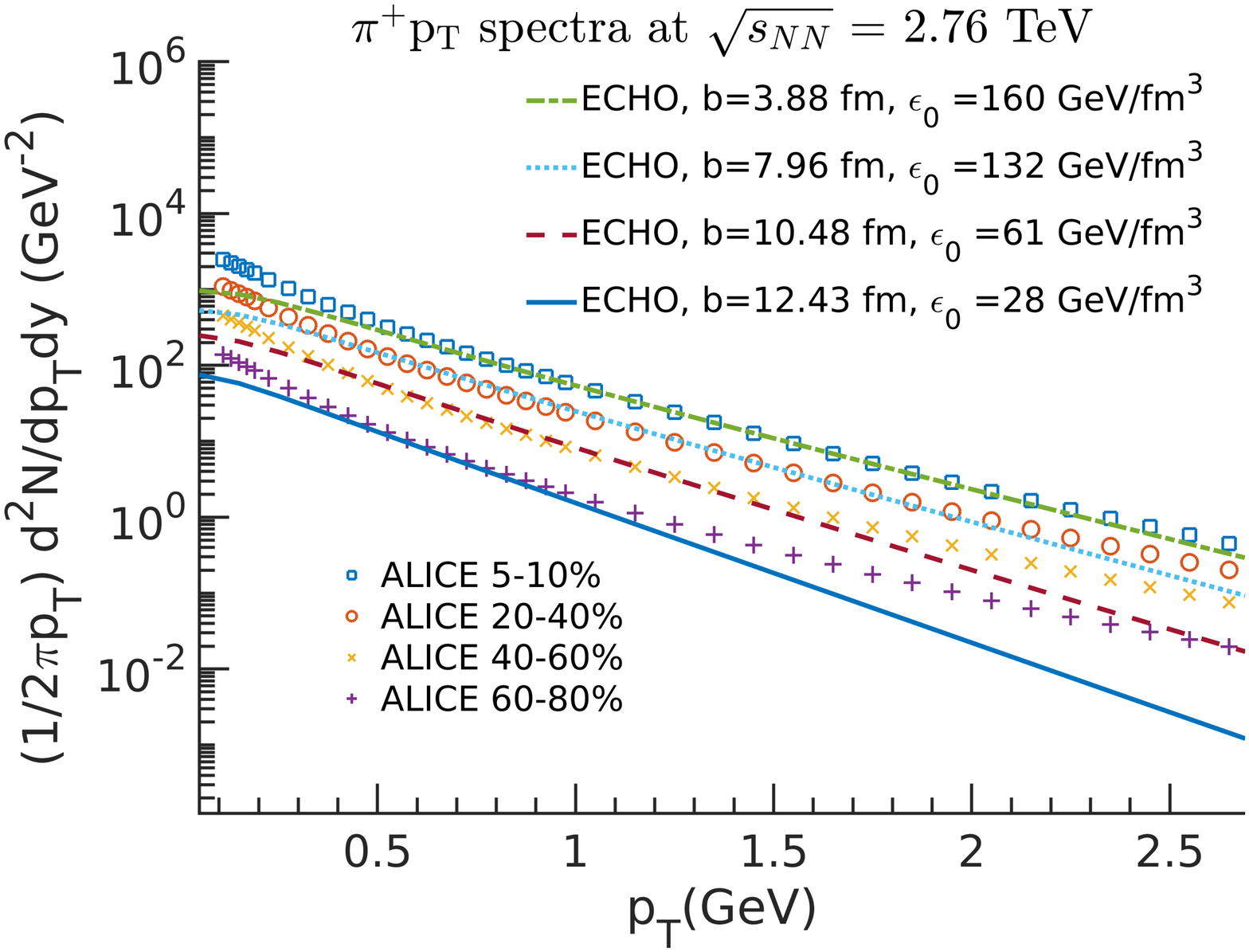}\\
\includegraphics[scale=0.34]{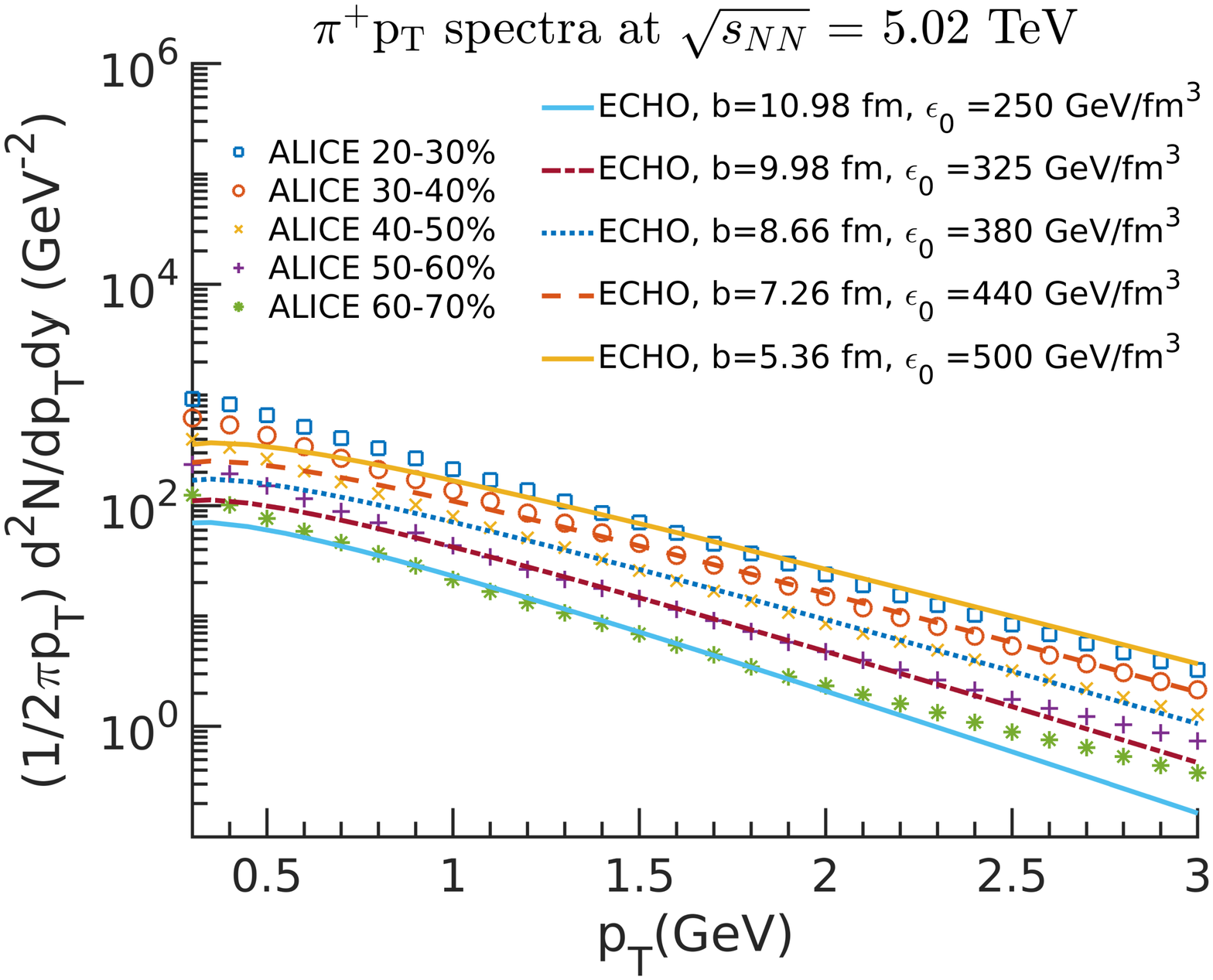}\\
\caption{Pion($\pi^+$) $p_T$ spectra for the two mentioned LHC energies and centralities compared with those obtained from ECHO-QGP.}
\label{fig:pT_spectra}
\end{center}
\end{figure}

The equation of state (EOS) from Wuppertal-Budapest(WB) Collaboration~\cite{WBEoS2010} replaces the earlier quasiparticle EOS~\cite{qpm2010}.
WB EOS is spline interpolation with the hadron Resonance Gas (HRG) EOS~\cite{hrg2010} for the hot and dense hadronic matter after hadronization. 
WB EOS computed from the lattice QCD is a better choice as the pseudo critical temperature range of QGP predicted by their analysis lie close to the presently agreed upon value~\cite{ding2020new}.   
We ran the hydrodynamics code for $11$ values of the impact parameter covering $0-100$\% centrality range. We chose the geometric Glauber initialization in ECHO-QGP~\cite{loizides2018improved}. A rapidity profile of p$-$p collisions is also employed as an input. The two parameters characterizing this profile are $\Delta_s$; which is the extension of the rapidity plateau and $\sigma_{\eta}$; which is the width of the  Gaussian falloff of the profile. Values for both of these parameters are varied until the shape of pions ($\pi^+$) rapidity spectra matches with that from the experimental poins rapidity spectra for the two mentioned LHC energies as shown in Fig(\ref{fig:rap_spectra})~\cite{2p76rap_spectra2013,5p02rap_spectra2017}.  
The values of the relaxation time coefficient for viscosity of second order, $\tau_{\pi}$ and the shear viscosity to entropy density ratio, $\eta/s$ are taken from ~\cite{busza2018heavy,song2010interplay}.
The thermalization time in the code is set for both the energies viz. $2.76$ TeV and $5.02$ TeV at $0.20$ fm/c~\cite{busza2018heavy,bhaduri2019anisotropic,krouppa2019bottomonium}.
The inelastic nucleon-nucleon cross-section is taken to be $61.8$ mb and $70$ mb for $2.76$ TeV and $5.02$ TeV, respectively~\cite{loizides2018improved}.
Lattice QCD predicts formation of a thermalized medium at energy density above $1.0$ GeV/$\mathrm{fm^3}$~\cite{KarschLQCD2002}. Initial energy density which goes as an input in ECHO-QGP was at first calculated roughly using an approximate relation $\epsilon_0 = \frac{1}{A_T \tau_0} J(y,\eta)\frac{dE_T}{dy}$ involving overlap area, initial thermalization time and the differential transverse energy~\cite{adler2005systematic,bjorken1983highly}.
But the peak values of the pion $p_T$-spectra for these values fall shorter than the experimental values. Hence, we varied the initial energy density at each centrality such that the pions $p_T$-spectra from ECHO-QGP matches with that from experiment values~\cite{5p02pT_spectra2020,2p76pT_spectra2017}.
The comparison of the pion spectras at few of these centralities is shown in Fig(\ref{fig:pT_spectra}).
The key parameters used in the ECHO-QGP hydrodynamics are summarized in Table~\ref{tab:table1}.

\begin{table}
\caption{ 
\label{tab:table1}
The Key parameters used as input in ECHO-QGP.-}
\begin{ruledtabular}
\begin{tabular}{l|cc}

\multicolumn{1}{c|}{\textbf{Parameters}} & \multicolumn{2}{c}{\textbf{Values}} \\ \hline
Initialization 	  & \multicolumn{2}{c}{Geometric Glauber} \\
Equation of State & \multicolumn{2}{c}{WB EoS spline interpolated with HRG} \\
Grid Size (fm)    & \multicolumn{2}{c}{30$\times$30$\times$30} \\
Grid points & \multicolumn{2}{c}{125} \\ [1ex]
\begin{tabular}[c]{@{}l@{}}Relaxation time for\\ [-0.4ex] viscosity, $\tau_{\pi}$\end{tabular} & \multicolumn{2}{c}{3.0} \\ [1.2ex]
$\eta/s$ & \multicolumn{2}{c}{0.1 $\approx$ 1.25 $\times (1/4\pi)$} \\ [1ex] \cline{2-3} 
 & \multicolumn{1}{c}{\textbf{\uline{For 2.76 TeV}}} & \textbf{\uline{For 5.02 TeV}} \\ 
 
Extension of the rapidity \\ plateau in p-p collision, $\Delta_s$  & \multicolumn{1}{c}{4} & 3.4   \\ [1.2ex]  

Width of the gaussian \\ falloff in pp collision , $\sigma_{\eta}$  & \multicolumn{1}{c}{1.5} & 1.7 \\ [1.2ex]

$t_{start}$(fm/c)      & \multicolumn{1}{c}{0.2} & 0.2 \\ 
$\sigma_{NN}$(mb)	      & \multicolumn{1}{c}{61.8}   & 70  \\ 
\end{tabular}

\end{ruledtabular}
\end{table}

The suppression formalism, described in the next section, requires temperature of the medium at different centralities and rapidities as a function of proper time in the transverse plane, which are obtained from the ECHO-QGP. Calculation of suppression at all transverse (x,y) points is computationally infeasible. Hence, temperatures are integrated over the transverse plane at all centrality, rapidity and proper time with a Gaussian weight factor and are taken as an input in the suppression formalism. The standard deviation of this Gaussian profile is varied within a specific range which is explained in the results and discussion section.

\subsection{Suppression formalism}
Bottom quark anti-quark pair produced in the hard scattering of colliding nuclei in the early stage of collision combines to form the bottomonia mesons. These heavy flavor mesons will drift in the medium and their decay products are eventually detected. 
During their time in the medium, these meson bound states are affected by various medium dependant dissociation mechanisms like color screening, collisional damping and gluonic dissociation, which are individually explained below along with the possible recombination due to correlated $b-\bar b$ pairs and the non-medium effect of shadowing.  

\subsubsection*{Gluonic Dissociation}
Gluonic dissociation is referred to the process where a bottomonium color singlet state absorbs a soft gluon in the medium and gets excited to a color octet state. The cross-section for this process is calculated as~\cite{Nendzig2013};
\begin{multline}
\sigma_{\mathrm{diss},nl}(E_g) = \\
 \frac{\pi^2 \alpha_s^u E_g}{N_c^2} \sqrt {\frac{m_q}{E_g+E_{nl}}} \frac{(l+1) |J^{q,l+1}_{nl}|^2 +  l | J^{q,l-1}_{nl}|^2}{2l+1}
 \label{eq:two}
\end{multline}
where, $E_g$ is the soft gluon energy, $E_{nl}$ is the eigenvalues corresponding to the bottomonium wavefunction ($g_{nl}(r)$), $m_q$ is bottom quark mass in GeV, $N_c$ is the number of color charges, $\alpha_s^u = \alpha_s(m_q\alpha_s^2 /2) \approx 0.59$.

\begin{equation}
J^{q,l^\prime}_{nl} = \int^\infty_0 r \; g^\ast_{nl}(r) \; h_{qi^\prime}(r) \; dr
\end{equation}
Above equation gives the probability density, where $g^\ast_{nl}$ and $h_{qi^\prime}(r)$ are the singlet and octet wavefunction of bottomonium respectively, obtained after numerically solving the $3$-dimensional Schr\"{o}dinger’s equation.
We integrate the cross-section in equation(\ref{eq:two}) with Bose-Einstein distribution as weight factor over gluon momentum to calculate the dissociation width due to gluonic dissociation i.e., $\Gamma_{\mathrm{gd},nl}$. 
The validity of the above cross-section assumes T$\ll1/r$, where T is the medium temperature and $r$ is the distance between quark and antiquark\cite{brambilla2011_gluo}. This regime is different than the one for which collisional damping holds, due to which our final results for T$\ll1/r$ may have less accuracy.

\subsubsection*{Collisional Damping}
We expect bottomonium to dissociate while it traverses through the plasma, due to the \textit{momentum transfer} arising out of collision. To account for this effect, we use a potential non-relativistic QCD (pNRQCD) formalism which depends on the imaginary part of the color potential between quark anti-quark pair. The complex potential between quark anti-quark pair located inside the QGP medium as determined by Laine et al.,~\cite{Laine2007} using effective field theory, is given as;

\begin{multline}
V(r,m_D) = \frac{\sigma}{m_D}(1 - e^{-m_Dr})  - \alpha_{\mathrm{eff}} \bigg( m_D + \frac{e^{-m_Dr}}{r} \bigg) \\  
- i \alpha_{\mathrm{eff}}T \int_0^{\infty} \frac{dz\;2z}{(1+z^2)^2} \bigg( 1 - \frac{sin(m_Drz)}{m_Drz} \bigg)
\end{multline}

Where, \\
$\alpha_{\mathrm{eff}} = 4 \alpha_s^s / 3$, \\ 
$\sigma$ is the string tension, whose value is $0.192$ $\mathrm{GeV^2}$, \\
$m_D$ is the Debye mass which is expressed as;
\begin{center}
$m_D = T \sqrt{4 \pi \alpha_s^T \bigg( \frac{N_c}{3} + \frac{N_f}{6} \bigg) }$. \\
\end{center}
Here, $N_c$ and $N_f$ are the number of color charges and number of flavors, respectively.\\
This potential is valid in the regime T$\gg 1/r \gtrsim m_D$~\cite{brambilla2013_damping}. 
We take the expectation value of the imaginary part of this potential to get the dissociation width corresponding to collisional damping~\cite{strickland2011thermal} as;

\begin{equation}
\Gamma_{\mathrm{damp},nl}(\tau,p_T,b)=\int g_{nl}(r)^\dagger \; \mathrm{Im}(V) \; g_{nl}(r) \; dr,
 \label{eq:cd}
\end{equation}
where, $g_{nl}(r)$ is the singlet wavefunction of the bottomonium.\\
In our previous work~\cite{singh2019}, the explicit $\tau$ dependence of gluonic dissociation widths and collisional damping widths(\ref{eq:cd}) arose from the analytical expression for temperature, $T(\tau)$, obtained by solving Bjorken's dynamics. We have now replaced it by the tabulated temperature values from ECHO-QGP at each centrality and rapidity integrated over the transverse plane for each of the bottomonium states.

\subsubsection*{Color Screening}

Due to the presence of free color charges in QGP medium, the bottom quark and anti-quark ($b and \bar{b}$) experiences a short range Yukawa-type color charge potential dependent on the medium temperature. As a consequence of this the formation of bound states in the medium is suppressed if the medium temperature goes beyond a certain temperature (dissociation temperature). This phenomenon is called the color screening~\cite{matsui1986} in an analogous way to the Debye charge screening in the quantum electrodynamics plasma. Different bottomonia species take different time, after collision, to form the corresponding bottom anti-bottom bound states termed as their respective formation times($\tau_f$). Consider a screening region in the fireball as a sphere with a screening radius($r_s$) defined for a bottomonium state depending on its dissociation temperature($T_D$). Suppose a $b\bar{b}$ quark pair forms at the position $\vec r_Q$, then it will likely form a bound state if it escapes the screening region in time equal to its formation time. The condition for which could be given by $|\vec r_Q+\vec v_T \tau_F|\geq r_s$, where $\vec v_T$ is the transverse drift velocity of the heavy quark in the medium. 
Here $\tau_F$ is the vacuum formation time of bottomonium.    

Considering the transverse motion of this bottomonium state, we can simplify the above condition as;
\begin{equation}
cos(\phi) \geq Y \quad; \mathrm{where},  Y = \frac{ ((r^2_s - r^2_Q)m_Q - \tau_f p^2_T / m_Q) } {2r_Q\tau_f p_T}
\label{eq:cs_condition}
\end{equation}

Here, $m_Q$ is the mass of the bottomonium state under consideration. Let us now consider a transverse radial distribution of $b\bar{b}$ produced after the hard scattering as:

\begin{equation}
h(r) = \Bigg( 1 - \frac{r^2}{\mathrm{R^2_T}} \Bigg) \theta(\mathrm{R_T}-r).
\end{equation}

Where $R_T$ is the maximum fireball radius obtained for different centralities from Modified Glauber analysis~\cite{loizides2019erratum}.

The color screening survival probability is given by:

\begin{equation}
{S_{cs}(p_T)} = \frac{4}{\pi R^2_T} \int_0^{R_T} dr \,r \,\phi_{max}(r) \Bigg( 1 - \frac{r^2}{\mathrm{R^2_T}} \Bigg)
\end{equation}

Where, $\phi_{max}(r)$ is the maximum positive azimuthal angle allowed by the condition in Eq(\ref{eq:cs_condition}).

\begin{figure}
\includegraphics[scale=0.52]{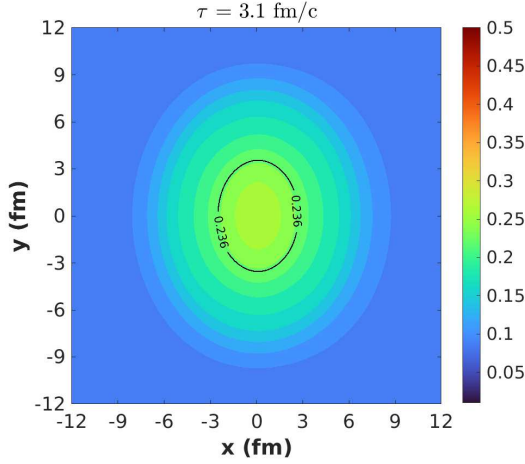} \\
\caption{The transverse temperature profile for 2.76 TeV collisional energy in ECHO-QGP at $N_{part}$ = $130$, rapidity = $1.44$ at $\tau$ = $3.1$ fm/c which is taken as the formation time of $\Upsilon(3S)$. The region inside the contour of $T_D=236$ MeV for $\Upsilon(3S)$ state depicts the screening region. As it is not for the most central collision, the fireball cross section is oval shaped.}
\label{fig:cs}
\end{figure}

In our previous work~\cite{singh2019}, we equated an assumed pressure profile in the transverse plane with the cooling law to obtain the screening time (time to drop initial pressure to the pressure corresponding to the dissociation temperature). Screening time is then equated to the bound state formation time (at the boundary of the screening region) to obtain the screening radii~\cite{MishraCS2007}.
Here we eliminated the need of this assumption of pressure profile in the transverse plane by directly finding the screening radii from the transverse temperature evolution using ECHO-QGP.
We take the minimum radius of the oval shaped contour shown in the Fig(\ref{fig:cs}) as the screening radius at the evolution time equal to the formation time for a given bottomonium species. The temperature contours are marked for the dissociation temperature of all the bottomonium states at all centralities.
The dissociation temperature ($T_D$) of different bottomonium states are borrowed from the analysis given in~\cite{DissoTemp2018}.

\subsubsection*{Shadowing}
The shadowing correction to $R_{AA}$ applied in our formalism is a modified version of the similar work by R. Vogt~\cite{vogt2010cold}.
We have replaced the shadowing factors used for gluons from {\it EPS09}~\cite{eskola2009eps09} to the more recent {\it EPPS16}~\cite{eskola2017epps16}. The central fit set is selected out of various available error sets in {\it EPPS16}. The parton distribution function of gluons have been updated to {\it CT14}~\cite{dulat2016new} from the earlier PDFs {\it CTEQ6}~\cite{pumplin2002new}. The contribution of suppression arising due to the shadowing effect is expressed as~\cite{armesto2006nuclear};
\begin{equation}
S_{sh}(p_T,b)= \frac{d\sigma_{AA}/dy}{T_{AA} d\sigma_{pp}/dy}.
\end{equation}
Shadowing effect influences the number of initially produced bottomonia($N_{Q}$). Hence, the shadowing corrected initial number of bottomonium is calculated as,$N^i_{Q}(\tau_0,b)=N_{Q}(\tau_0,b) S_{sh}(p_T,b)$.

\subsection{Recombination mechanisms}
We have incorporated the possibility of recombination of $b\bar{b}$ due to de-excitation from octet to singlet state with a gluon emission, even though it will be negligible for the case of bottomonium. We find the recombination cross-section in QGP using detailed balance from gluonic dissociation cross-section as~\cite{thews2006momentum}
\begin{equation}
\sigma_{f,nl} = \frac{48}{36}\sigma_{d,nl}\frac{(s-M^2_{nl})^2}{s(s-4m_q m_{\bar{q}})},
\end{equation}
where, $s$ is the Mandelstan variable, $M_{nl}$, $m_q$ and $m_{\bar{q}}$ are the masses of bottomonia states, bottom quark and bottom anti-quark, respectively. 
We then define a recombination factor as the thermal average of the product of the above cross-section and relative velocity between $b$ and $\bar{b}$ as, $\Gamma_{F,nl} = \langle \sigma_{f,nl} \; v_{rel} \rangle_k$.

\subsection{Final Number of Bottomonium}
Due to all of the above effects, the bottomonia can dissociate or the correlated $b\bar{b}$ pair can recombine again into bound states. 
We assume that this interplay of dissociation and recombination is governed by a simple first order differential equation given as~\cite{thews};
\begin{equation}
\frac{dN_Q(\tau)}{d\tau} = \frac{\Gamma_{F,nl}N_q N_{\bar{q}}}{V(\tau)} - \Gamma_{D,nl} N_{Q} 
\end{equation}
Here, $N_{Q}$ is the bottomonia yield at a given value of proper time ($\tau$). First and second terms on the right hand side of this equation correspond to the recombination and dissociation terms, respectively. $\Gamma_{F,nl}$ and $\Gamma_{D,nl}$ are the corresponding recombination and dissociation rates. $N_q$ and $N_{\bar{q}}$ are the number of heavy quark and anti-quark produced in p$-$p collision. $V(\tau)$ is the instantaneous volume of the expanding fireball. 

The solution for the above first order differential equation under the approximation that, $N_Q<N_{q}$,$N_{\bar{q}}$ is given by,
\begin{align}
N_{Q}(\tau_{QGP},p_{T}&) = \epsilon(\tau_{QGP},p_T) \bigg[ N_{Q}(\tau_0)\nonumber\\
&+ N_q N_{\bar{q}} \int^{\tau_{QGP}}_{\tau_0} \frac{ 
\Gamma_{F,nl}(\tau,p_T)}{V(\tau) \; \epsilon(\tau,p_T)} d\tau \bigg],
\label{eq:one}
\end{align}
where $\tau_{QGP}$ is the QGP lifetime and $\tau_0$ is the initial time at which we start hydrodynamics and which also marks the beginning of the QGP stage. 

The first term inside the bracket on the right hand side in Equation (\ref{eq:one}) is the bottomonia produced at the initial hard scattering stage. $N_{Q}(\tau_0)$ is calculated as~\cite{miller2007glauber};
\begin{equation}
N_{Q}(\tau_0,b)=\sigma^{NN}_{Q} T_{AA}(b),
\end{equation}
where, $\sigma^{NN}_{Q}$ is the production cross-section of the bottomonium at a given collisional energy. $T_{AA}(b)$ is the nuclear overlap function.  
$V(\tau)$ in equation (\ref{eq:one}) is volume of fireball given by the formula;
\begin{equation}
V(\tau) = \tau_0\pi R^2_T\bigg(\frac{\tau_0}{\tau}\bigg)^{\frac{1}{R}-1}
\end{equation}

where, $R_T$ is the radius of colliding nuclei and $R$ is the Reynold’s number.\\
$\epsilon(\tau,p_T)$ in equation (\ref{eq:one}) is a suppression factor integrated cumulatively in $\tau$ values and is given by;
\begin{equation}
\epsilon(\tau,p_T) = \exp \bigg(-\int^{\tau}_{\tau^{\prime}_{nl}}\Gamma_{D,nl}(\tau^{\prime},p_T) d\tau^{\prime}\bigg).
\end{equation}

Equation (\ref{eq:one}) also has multiplicative suppression factor which is integrated for the complete QGP lifetime and is calculated as;

\begin{equation}
\epsilon(\tau_{QGP},p_T) = exp \bigg( -\int^{\tau_{QGP}}_{\tau^{\prime}_{nl}} \Gamma_{D,nl}(\tau,p_T) d\tau \bigg)
\end{equation}

In the above equation, $\tau^{\prime}_{nl}$ is the time required for the formation of a given bottomonium state within QGP.

We then calculate the ratio, $S_p^\prime$ = $N_Q$/$N_Q^i$ referred to as the survival probability. The color screening has been considered as an independent effect and hence the total survival probability is calculated as $S_p = S_p^\prime \times S_{cs}$. We find this survival probability for 5 bottomonia states which are $\Upsilon(1S)$, $\Upsilon(2S)$, $\chi_{b0}(1P)$, $\chi_{b0}(2P)$ and $\Upsilon(3S)$. A feed-down scheme dictates the total yield after all possible decays from higher excited states~\cite{bottomganesh2013}. The Survival probability($S_p$) obtained after feed down for $\Upsilon(1S)$ and $\Upsilon(2S)$ is plotted and compared with the respective $R_{AA}$ obtained from experiments. 

\section{Results and Discussions}

\begin{figure*}
\begin{tabular}{cc}
\includegraphics[scale=0.34]{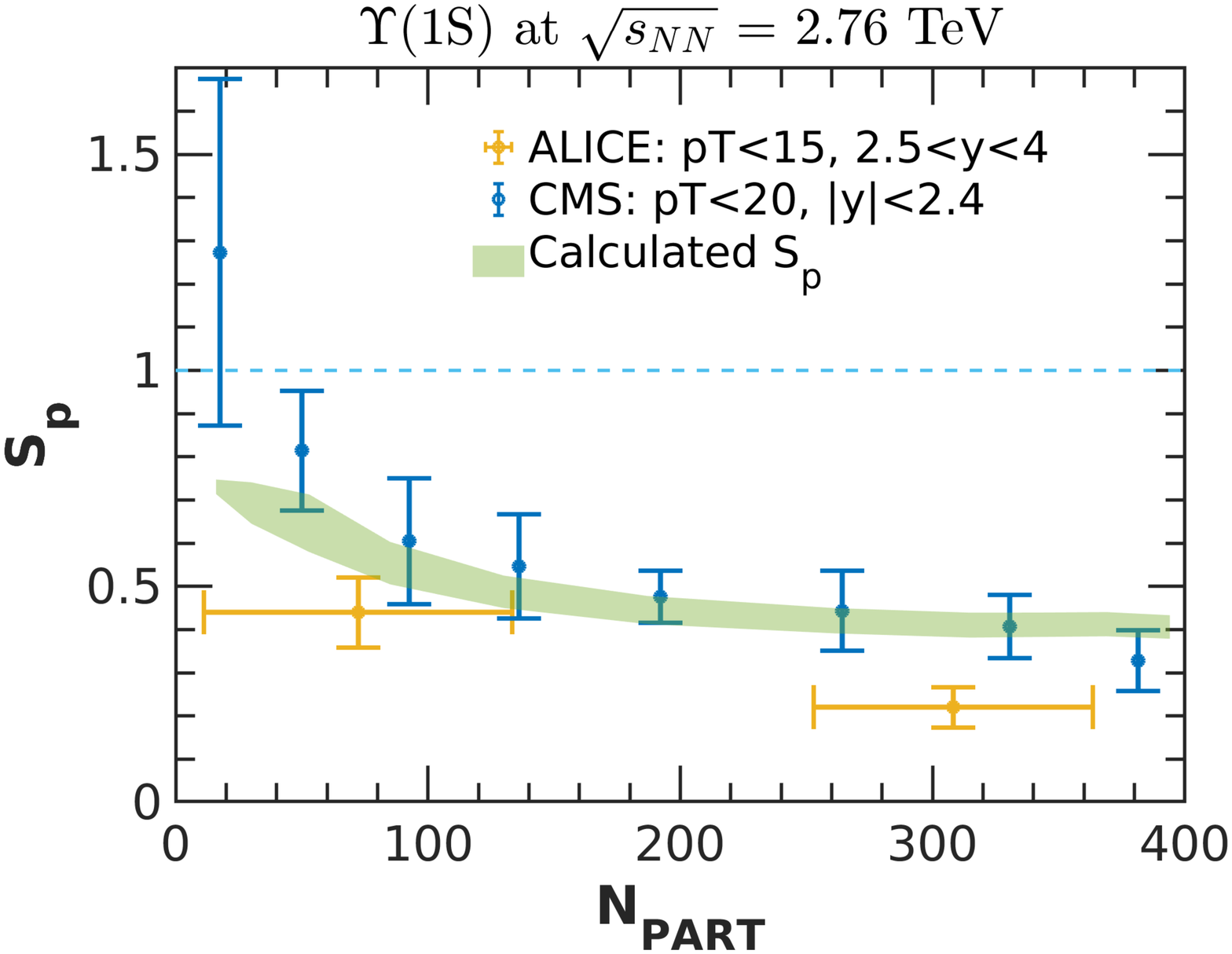}   &  \includegraphics[scale=0.34]{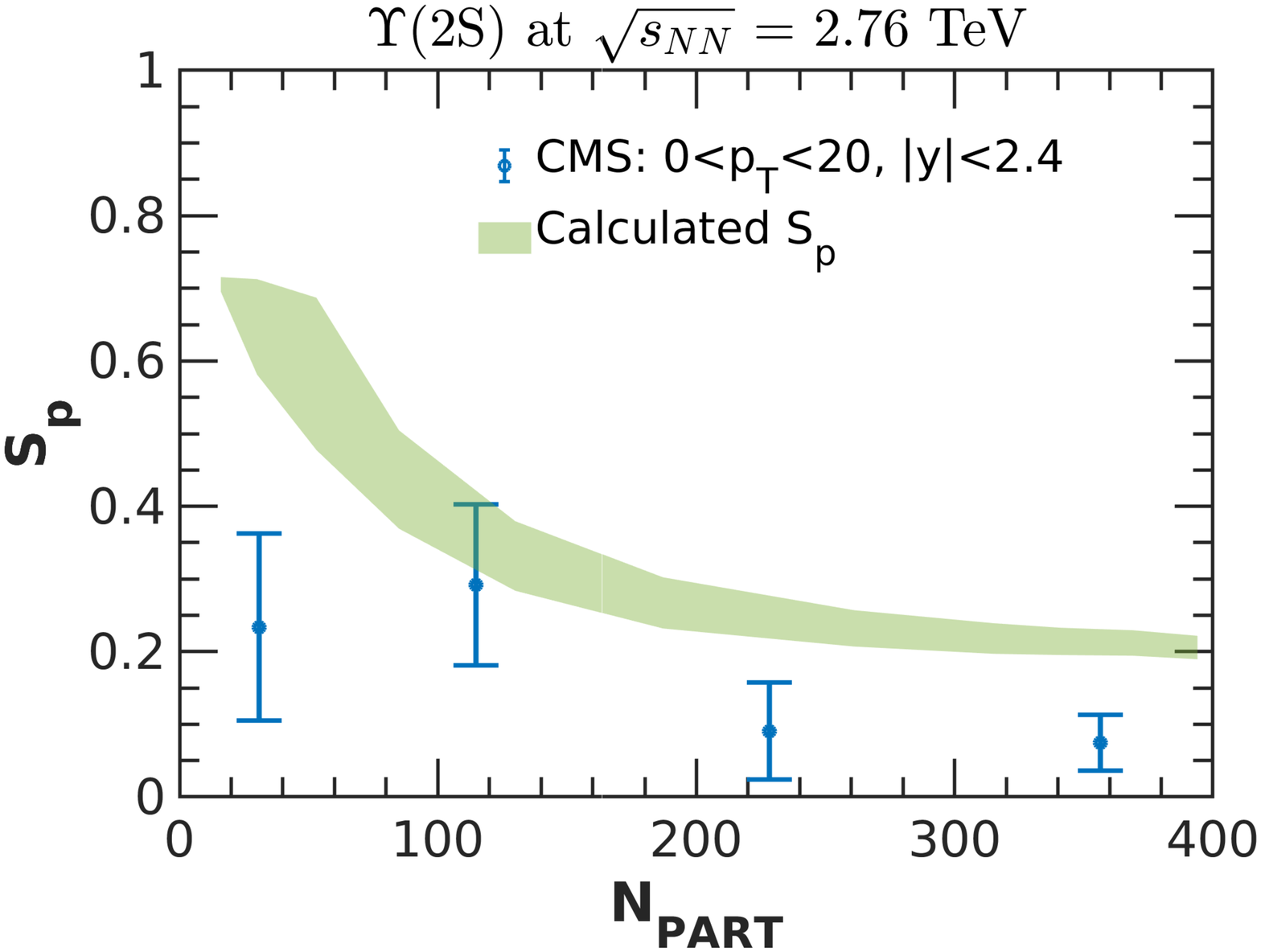} 
\end{tabular}
\caption{Centrality dependence of suppression for $\Upsilon$ compared with measured $R_{AA}$ at $\sqrt{s_{NN}} = 2.76$ TeV.}
\label{fig:npart_2p76}
\end{figure*}

\begin{figure*}
\begin{tabular}{cc}
\includegraphics[scale=0.34]{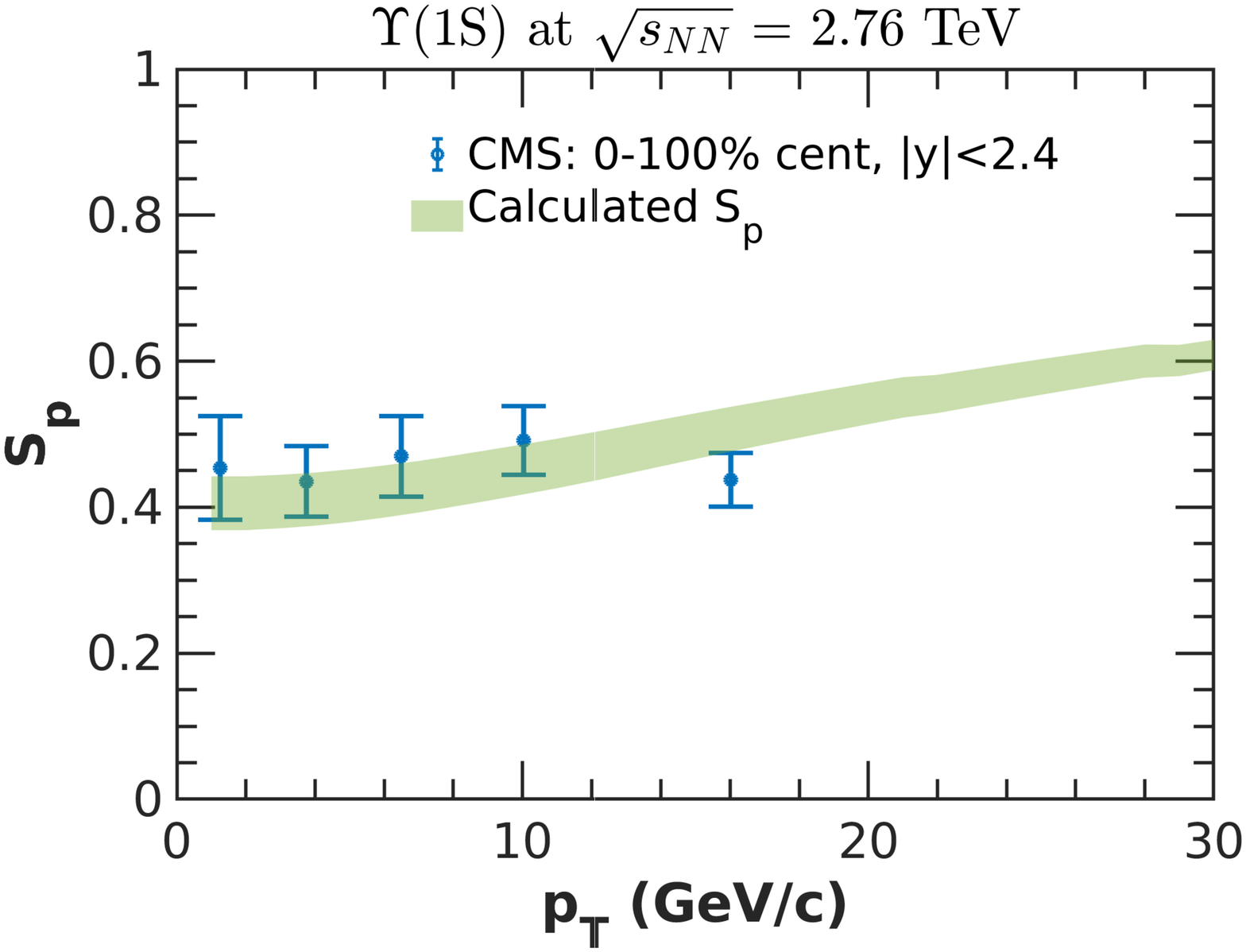}  &  \includegraphics[scale=0.34]{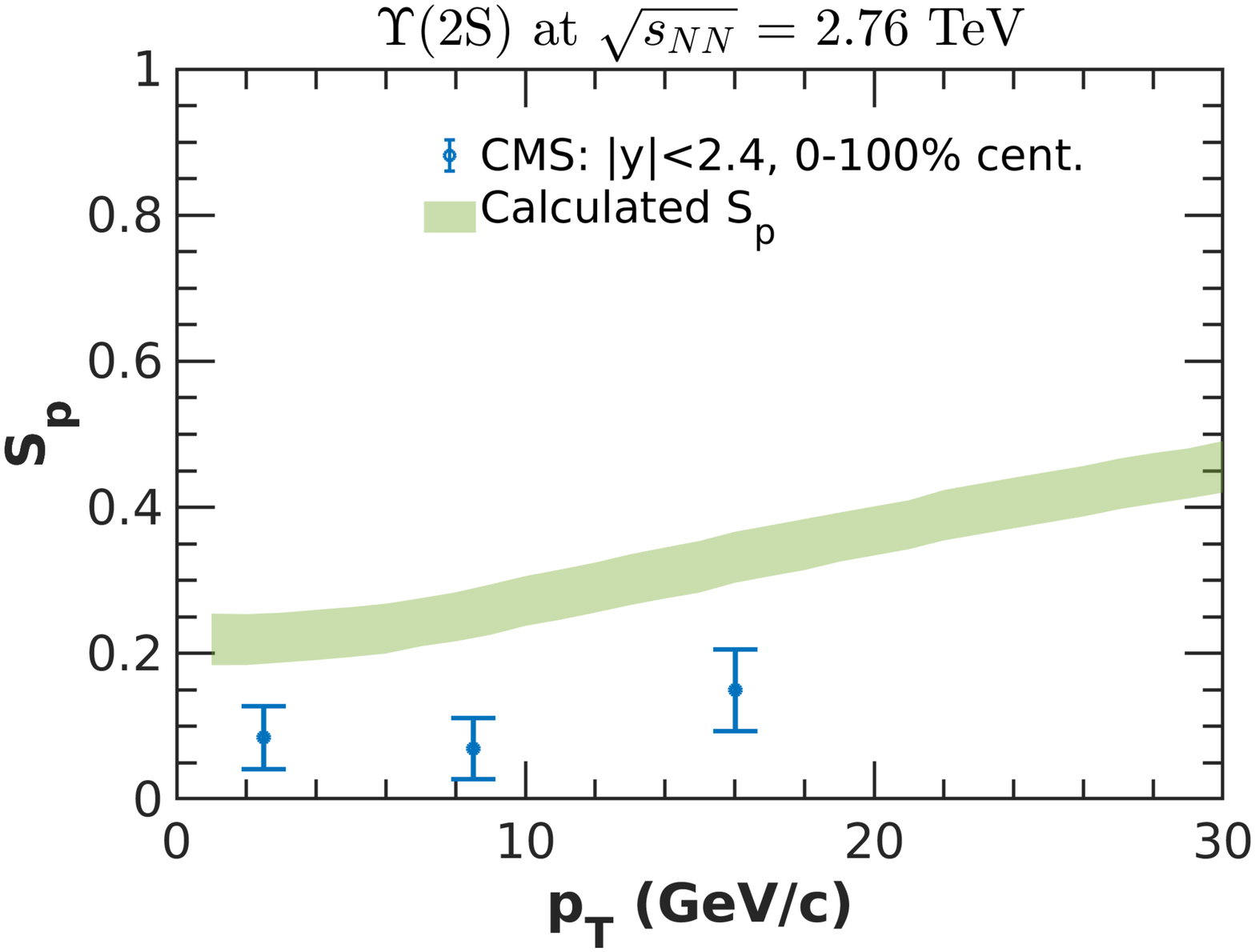}
\end{tabular}
\caption{Transverse momentum dependence of suppression for $\Upsilon$ compared with measured $R_{AA}$ at $\sqrt{s_{NN}}$ = $2.76$ TeV}
\label{fig:pT_2p76}
\end{figure*}

Results have been generated for centrality range of $0-100$\%, transverse momentum range of $1-30$ GeV/c and rapidity range of $|y| < 3.6$ which covers the ranges of experimentally available data from CMS and ALICE at both $2.76$ TeV and $5.02$ TeV~\cite{khachatryan2017suppression,alice2014suppression,sirunyan2019measurement,acharya2019Upsilon}.

ALICE suppression data at forward rapidity complements well with the broader rapidity range of CMS data especially for $p_T$ and $N_{part}$ dependence of $\Upsilon(2S)$ at $\sqrt{s_{NN}}=5.02$ TeV. The standard deviation ($\sigma_T$) of the Gaussian weight factor used for integrating temperatures from ECHO-QGP as mentioned in Section (II-A) has been varied to obtain an uncertainty patch in the theoretically calculated suppression results as shown in the figures.
For $2.76$ TeV the standard deviation value lies between $0.7 < \sigma_T < 2.8$ and for $5.02$ TeV, it spans as $1.4 < \sigma_T < 3.9$. Selecting $\sigma_T$ greater than the upper limits for the two energies makes the QGP lifetime, in peripheral collisions at extreme rapidity values, smaller than the formation time of $\Upsilon(1S)$ and $\Upsilon(2S)$ states. Selecting $\sigma_T$ smaller than the lower limits undermines the temperatures at large x-y values rendering the purpose of integration futile.

Fig.(\ref{fig:npart_2p76}) depicts the centrality dependence of suppression for $\Upsilon(1S)$ and $\Upsilon(2S)$ states at $2.76$ TeV as calculated by our present model. The corresponding experimental suppression data are shown for comparison. Our calculated values of the survival probability for $\Upsilon(1S)$ lie very close to the CMS data and are following the trend of ALICE data. Whereas for $\Upsilon(2S)$, our predicted values are slightly less suppressed but mostly following the CMS data. 

\begin{figure*}
\begin{tabular}{cc}
\includegraphics[scale=0.34]{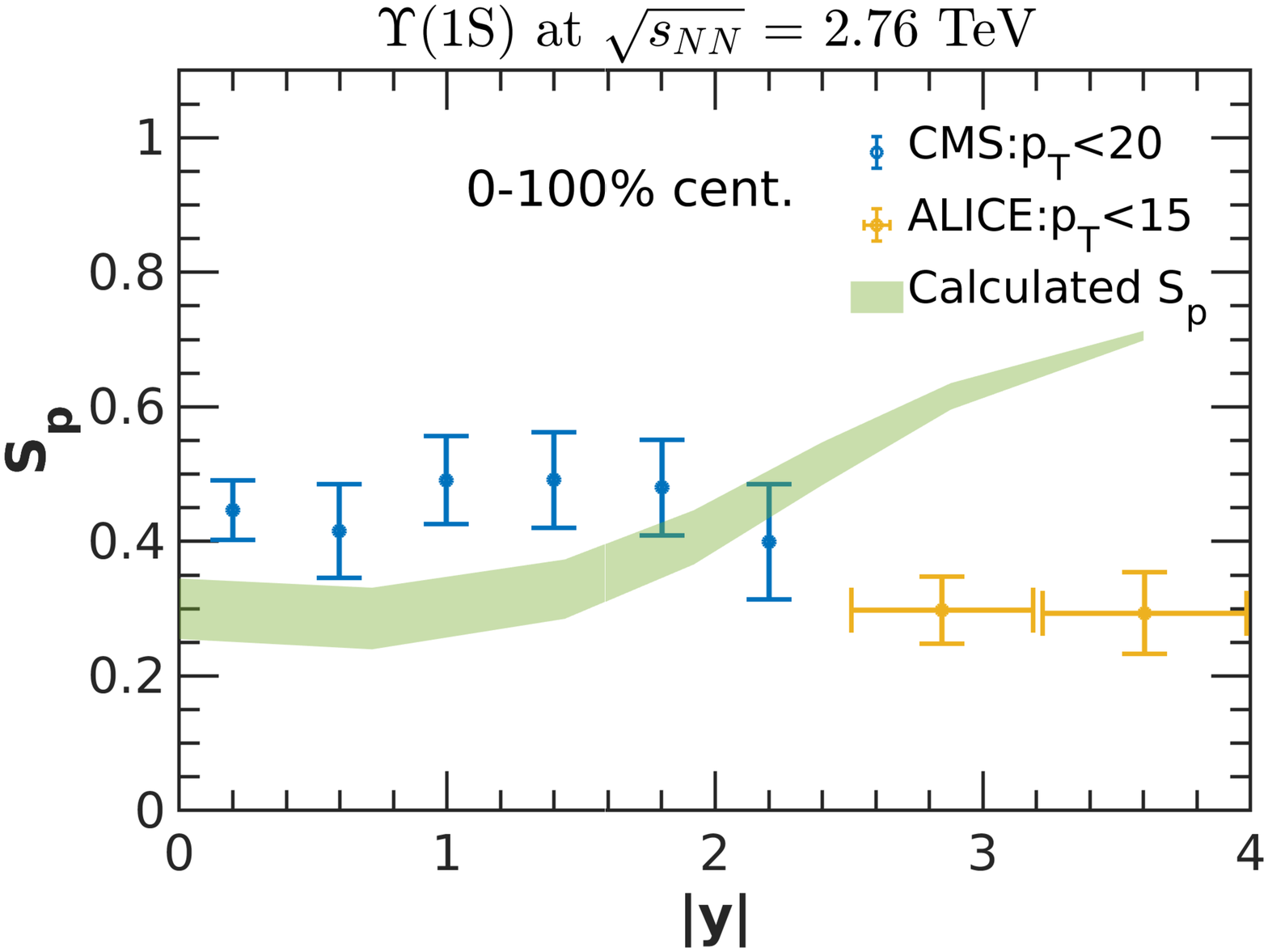}  &  \includegraphics[scale=0.34]{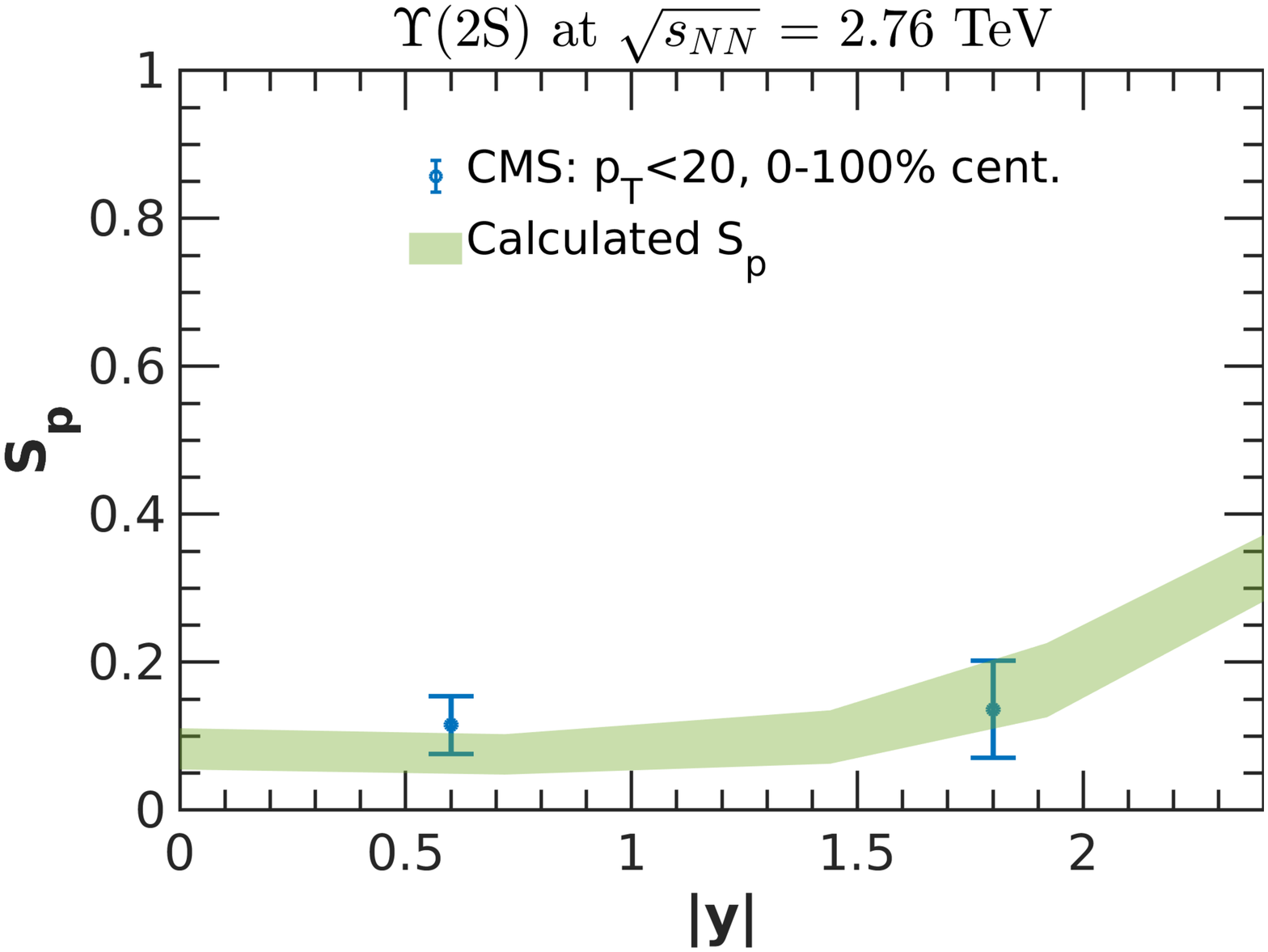}  
\end{tabular}
\caption{Rapidity dependence of suppression for $\Upsilon$ compared with $R_{AA}$ at $\sqrt{s_{NN}}$ = $2.76$ TeV.}
\label{fig:rap_2p76}
\end{figure*}

\begin{figure*}
\begin{tabular}{cc}
\includegraphics[scale=0.34]{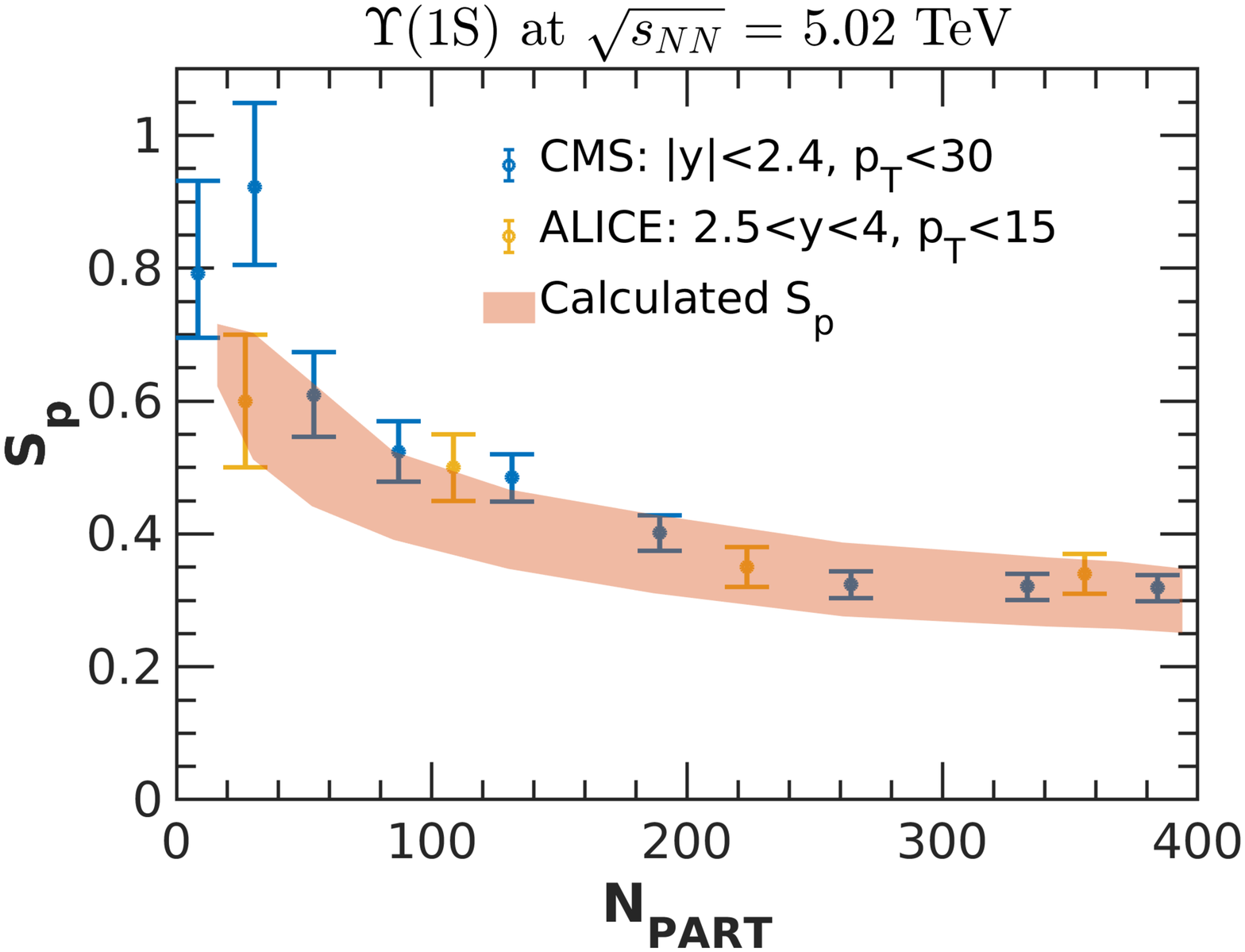} & \includegraphics[scale=0.34]{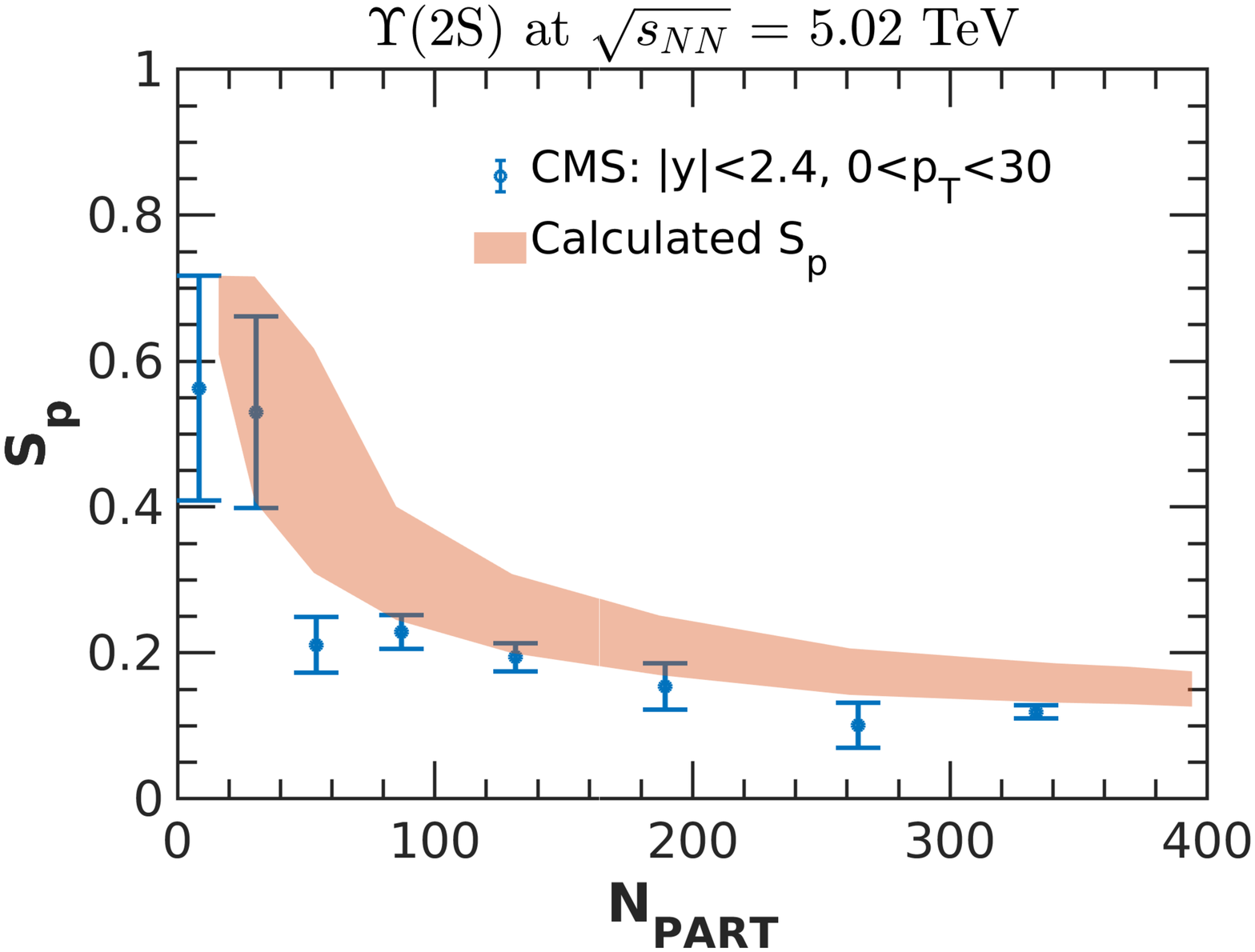}  
\end{tabular}
\caption{Centrality dependence of suppression for $\Upsilon$ compared with $R_{AA}$ at $\sqrt{s_{NN}}$ = $5.02$ TeV.}
\label{fig:npart_5p02}
\end{figure*}

\begin{figure*}
\begin{tabular}{cc}
\includegraphics[scale=0.34]{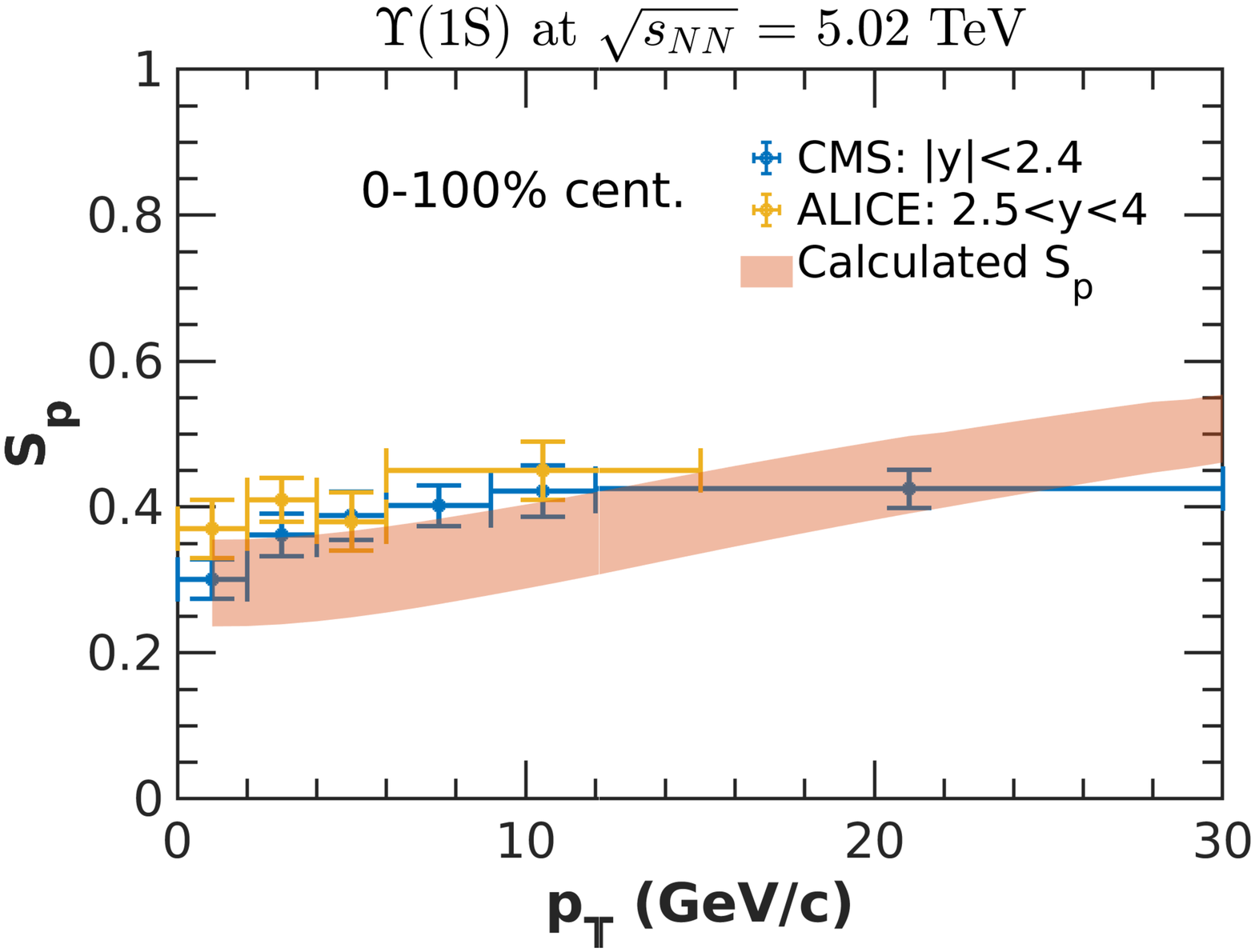}  & \includegraphics[scale=0.34]{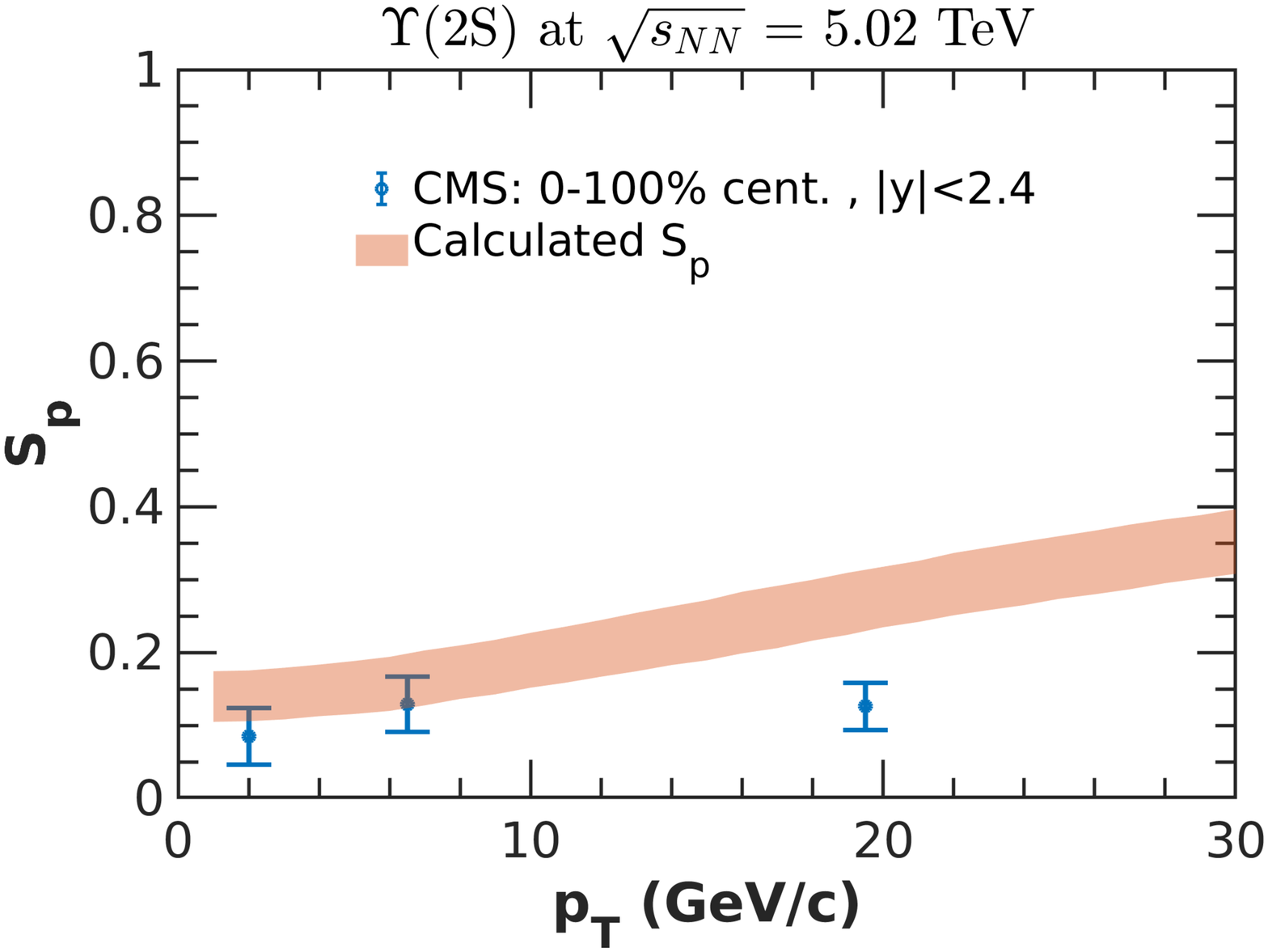}  
\end{tabular}
\caption{Transverse momentum dependence of suppression for $\Upsilon$ compared with $R_{AA}$ at $\sqrt{s_{NN}}$ = $5.02$ TeV.}
\label{fig:pT_5p02}
\end{figure*}

\begin{figure*}
\begin{tabular}{cc}
\includegraphics[scale=0.34]{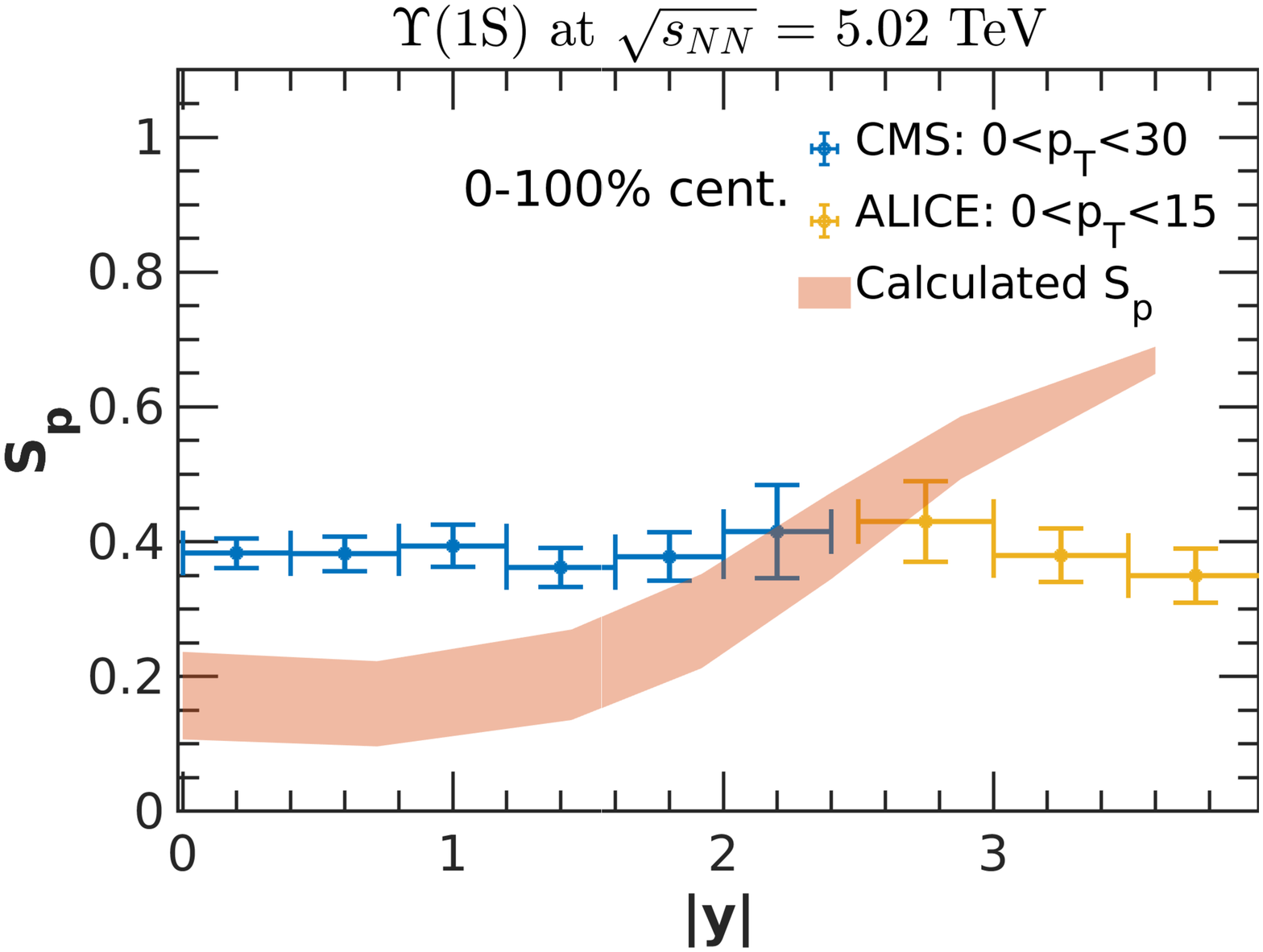}  &  \includegraphics[scale=0.34]{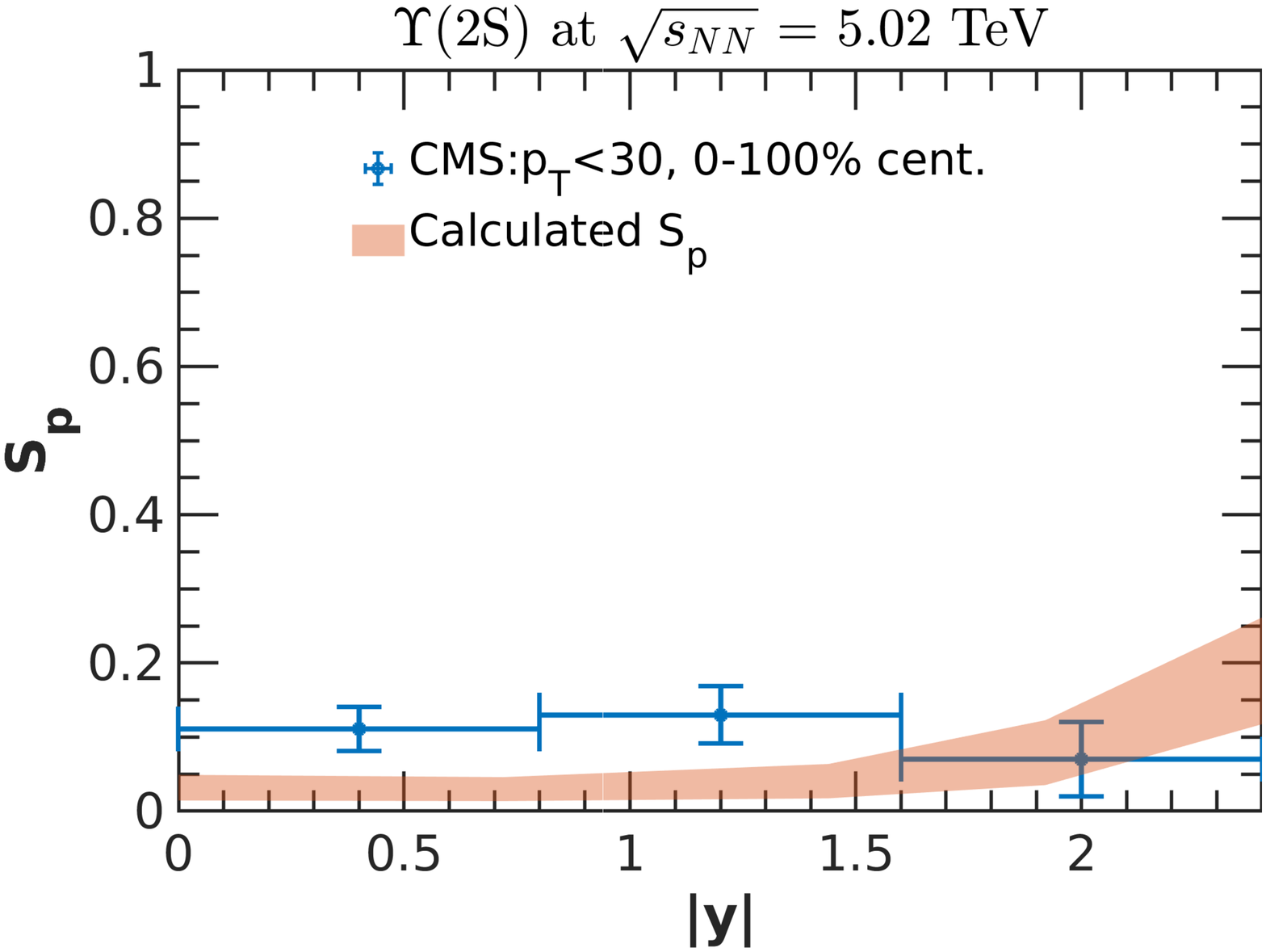}
\end{tabular}
\caption{Rapidity dependence of suppression for $\Upsilon$ compared with $R_{AA}$ at $\sqrt{s_{NN}}$ = $5.02$ TeV.}
\label{fig:rap_5p02}
\end{figure*}

Fig.(\ref{fig:pT_2p76}) shows the variation of our predicted values of suppression for $\Upsilon(1S)$ and $\Upsilon(2S)$ with respect to the transverse momentum at the $2.76$ TeV center of mass energy. We find that the agreement among our calculated and measured values for $\Upsilon(1S)$ are reasonably well. However, for $\Upsilon(2S)$, calculated suppression is slightly lesser than the values from available data from CMS.\\

In Fig.(\ref{fig:rap_2p76}), we have plotted our theoretical results of rapidity dependence of $\Upsilon(1S)$ and $\Upsilon(2S)$ suppression along with the corresponding experimental data. We find a quite reasonable agreement in the case of $\Upsilon(1S)$ than $\Upsilon(2S)$ within the uncertainty limit. Our predicted values show lesser suppression for $\Upsilon(2S)$ as compared with the corresponding experimental data. This rapidity dependence of suppression has become possible due to the interfacing of our earlier model with the ECHO-QGP's hydrodynamic expansion. Overall, for $2.76$ TeV LHC center of mass energy, we find good agreement between our calculated bottomonium suppression values and the corresponding experimentally available data under the theoretical and measured uncertainty limit.\\

Fig.(\ref{fig:npart_5p02}) depicts the variation of survival probability for both the bottomonium states with respect to the centrality at $5.02$ TeV center of mass energy. The measured values of suppression from ALICE and CMS data shown on the same plot for comparison for $\Upsilon(1S)$ associated with the centrality dependence are quite close to each other. We clearly see a quite good agreement from the Fig.(\ref{fig:npart_5p02}) between our predicted values and the measured ones for both the bottomonium states over the whole range of centrality. Since at $5.02$ TeV LHC energy.\\

Transverse momentum dependence of the survival probability values are plotted in Fig.(\ref{fig:pT_5p02}) along with the measured values of $R_{AA}$ for both the bottomonium states at $5.02$ TeV, center of mass energy. The agreement for $\Upsilon(1S)$ state is reasonably well especially for low $p_T$ values, whereas for $\Upsilon(2S)$, the predicted values are slightly less suppressed as compared to the CMS data.

Finally, rapidity dependent survival probability at $5.02$ TeV energy is shown in Fig.(\ref{fig:rap_5p02}) for both the bottomonium states. In comparison with the corresponding experimental data, our results are close to the measured values for $\Upsilon(2S)$. For $\Upsilon(1S)$, our calculated suppression is not following the trend of ALICE and CMS data. We saw similar disagreement for the corresponding rapidity results at 2.76 TeV energy. These results might improve with further refining of the input parameters taken in ECHO-QGP. 

Thus a complete dependence of bottomonium suppression for both the states spanning over two LHC energies has shown a convincing agreement with the measured values.   

\section{Summary and Conclusion}

To summarize, we have used a quarkonia suppression formalism to explain bottomonium suppression data at $2.76$ TeV and $5.02$ TeV LHC energies. ECHO-QGP has allowed us to find ($3+1$)-dimensional evolution of the relevant physical quantities associated with the medium formed just after the collisions. The temperatures at different centralities and rapidities have been extracted from the ECHO-QGP and fed into the suppression formalism. This facilitated us to include the rapidity dependence of suppression in the formalism. The experimental data on transverse momentum($p_T$) and rapidity($y$) spectra has been utilized to fix the input parameters of the hydrodynamics. We have used here the EOS from the Wuppertal-Budapest collaboration which is computed from first principle Lattice QCD. Modified color screening formalism has been used with the newer dynamics eliminating the need to assume a pressure profile for collisions in the transverse plane. Shadowing effect has been updated with the recent gluon PDFs and shadowing factors.\\
The uncertainty in temperature integration from ECHO-QGP has translated into an uncertainty patch in the final suppression results. This formalism has adequately explained the centrality and transverse momentum dependent suppression of $\Upsilon(1S)$ and $\Upsilon(2S)$ at both the LHC energies under consideration. Although, the calculated rapidity dependence of suppression does not follow the trend of data for $\Upsilon(1S)$, yet is quite satisfactory for $\Upsilon(2S)$ state at both the energies.  
Looking ahead, it will be interesting to do an open heavy flavor mesons evolution study in the QGP medium with a formalism suited well for such heavy-light mesons. A simultaneous study of open and hidden heavy flavor and will give a more reliable constraint on the model parameters since it will account for suppression for all heavy quarks bound states produced in the collisions.

\begin{acknowledgments}
We are grateful to Gabriele Inghirami for clearing our doubts in using ECHO-QGP from time to time. We would also like to thank Prashant Srivastava for discussions on various topics. M. Mishra is thankful to the Department of Science and Technology (DST), New Delhi for providing the financial assistance.
\end{acknowledgments}

\bibliography{1REFERENCES}

\providecommand{\noopsort}[1]{}\providecommand{\singleletter}[1]{#1}%
\begin{thebibliography}{65}%
\makeatletter
\providecommand \@ifxundefined [1]{%
 \@ifx{#1\undefined}
}%
\providecommand \@ifnum [1]{%
 \ifnum #1\expandafter \@firstoftwo
 \else \expandafter \@secondoftwo
 \fi
}%
\providecommand \@ifx [1]{%
 \ifx #1\expandafter \@firstoftwo
 \else \expandafter \@secondoftwo
 \fi
}%
\providecommand \natexlab [1]{#1}%
\providecommand \enquote  [1]{``#1''}%
\providecommand \bibnamefont  [1]{#1}%
\providecommand \bibfnamefont [1]{#1}%
\providecommand \citenamefont [1]{#1}%
\providecommand \href@noop [0]{\@secondoftwo}%
\providecommand \href [0]{\begingroup \@sanitize@url \@href}%
\providecommand \@href[1]{\@@startlink{#1}\@@href}%
\providecommand \@@href[1]{\endgroup#1\@@endlink}%
\providecommand \@sanitize@url [0]{\catcode `\\12\catcode `\$12\catcode
  `\&12\catcode `\#12\catcode `\^12\catcode `\_12\catcode `\%12\relax}%
\providecommand \@@startlink[1]{}%
\providecommand \@@endlink[0]{}%
\providecommand \url  [0]{\begingroup\@sanitize@url \@url }%
\providecommand \@url [1]{\endgroup\@href {#1}{\urlprefix }}%
\providecommand \urlprefix  [0]{URL }%
\providecommand \Eprint [0]{\href }%
\providecommand \doibase [0]{https://doi.org/}%
\providecommand \selectlanguage [0]{\@gobble}%
\providecommand \bibinfo  [0]{\@secondoftwo}%
\providecommand \bibfield  [0]{\@secondoftwo}%
\providecommand \translation [1]{[#1]}%
\providecommand \BibitemOpen [0]{}%
\providecommand \bibitemStop [0]{}%
\providecommand \bibitemNoStop [0]{.\EOS\space}%
\providecommand \EOS [0]{\spacefactor3000\relax}%
\providecommand \BibitemShut  [1]{\csname bibitem#1\endcsname}%
\let\auto@bib@innerbib\@empty
\bibitem [{\citenamefont {Jacak}\ and\ \citenamefont
  {M{\"u}ller}(2012)}]{jacak2012exploration}%
  \BibitemOpen
  \bibfield  {author} {\bibinfo {author} {\bibfnamefont {B.~V.}\ \bibnamefont
  {Jacak}}\ and\ \bibinfo {author} {\bibfnamefont {B.}~\bibnamefont
  {M{\"u}ller}},\ }\href {https://doi.org/10.1126/science.1215901} {\bibfield
  {journal} {\bibinfo  {journal} {Science}\ }\textbf {\bibinfo {volume}
  {337}},\ \bibinfo {pages} {310} (\bibinfo {year} {2012})}\BibitemShut
  {NoStop}%
\bibitem [{\citenamefont {Guenther}(2021)}]{guenther2021overview}%
  \BibitemOpen
  \bibfield  {author} {\bibinfo {author} {\bibfnamefont {J.~N.}\ \bibnamefont
  {Guenther}},\ }\href@noop {} {\bibfield  {journal} {\bibinfo  {journal} {The
  European Physical Journal A}\ }\textbf {\bibinfo {volume} {57}},\ \bibinfo
  {pages} {1} (\bibinfo {year} {2021})}\BibitemShut {NoStop}%
\bibitem [{\citenamefont {Adams}\ \emph {et~al.}(2005)\citenamefont {Adams},
  \citenamefont {Aggarwal}, \citenamefont {Ahammed}, \citenamefont {Amonett},
  \citenamefont {Anderson}, \citenamefont {Arkhipkin}, \citenamefont
  {Averichev}, \citenamefont {Badyal}, \citenamefont {Bai}, \citenamefont
  {Balewski} \emph {et~al.}}]{adams2005experimental}%
  \BibitemOpen
  \bibfield  {author} {\bibinfo {author} {\bibfnamefont {J.}~\bibnamefont
  {Adams}}, \bibinfo {author} {\bibfnamefont {M.}~\bibnamefont {Aggarwal}},
  \bibinfo {author} {\bibfnamefont {Z.}~\bibnamefont {Ahammed}}, \bibinfo
  {author} {\bibfnamefont {J.}~\bibnamefont {Amonett}}, \bibinfo {author}
  {\bibfnamefont {B.}~\bibnamefont {Anderson}}, \bibinfo {author}
  {\bibfnamefont {D.}~\bibnamefont {Arkhipkin}}, \bibinfo {author}
  {\bibfnamefont {G.}~\bibnamefont {Averichev}}, \bibinfo {author}
  {\bibfnamefont {S.}~\bibnamefont {Badyal}}, \bibinfo {author} {\bibfnamefont
  {Y.}~\bibnamefont {Bai}}, \bibinfo {author} {\bibfnamefont {J.}~\bibnamefont
  {Balewski}}, \emph {et~al.},\ }\href
  {https://doi.org/https://doi.org/10.1016/j.nuclphysa.2005.03.085} {\bibfield
  {journal} {\bibinfo  {journal} {Nuclear Physics A}\ }\textbf {\bibinfo
  {volume} {757}},\ \bibinfo {pages} {102} (\bibinfo {year}
  {2005})}\BibitemShut {NoStop}%
\bibitem [{\citenamefont {Krzewicki}\ \emph {et~al.}(2011)\citenamefont
  {Krzewicki}, \citenamefont {Collaboration} \emph
  {et~al.}}]{krzewicki2011elliptic}%
  \BibitemOpen
  \bibfield  {author} {\bibinfo {author} {\bibfnamefont {M.}~\bibnamefont
  {Krzewicki}}, \bibinfo {author} {\bibfnamefont {A.}~\bibnamefont
  {Collaboration}}, \emph {et~al.},\ }\href
  {https://doi.org/http://dx.doi.org/10.1088/0954-3899/38/12/124047} {\bibfield
   {journal} {\bibinfo  {journal} {Journal of Physics G: Nuclear and Particle
  Physics}\ }\textbf {\bibinfo {volume} {38}},\ \bibinfo {pages} {124047}
  (\bibinfo {year} {2011})}\BibitemShut {NoStop}%
\bibitem [{\citenamefont {Kovtun}\ \emph {et~al.}(2005)\citenamefont {Kovtun},
  \citenamefont {Son},\ and\ \citenamefont {Starinets}}]{kovtun2005viscosity}%
  \BibitemOpen
  \bibfield  {author} {\bibinfo {author} {\bibfnamefont {P.~K.}\ \bibnamefont
  {Kovtun}}, \bibinfo {author} {\bibfnamefont {D.~T.}\ \bibnamefont {Son}},\
  and\ \bibinfo {author} {\bibfnamefont {A.~O.}\ \bibnamefont {Starinets}},\
  }\href {https://doi.org/10.1103/PhysRevLett.94.111601} {\bibfield  {journal}
  {\bibinfo  {journal} {Physical review letters}\ }\textbf {\bibinfo {volume}
  {94}},\ \bibinfo {pages} {111601} (\bibinfo {year} {2005})}\BibitemShut
  {NoStop}%
\bibitem [{\citenamefont {Ratti}(2018)}]{ratti2018lattice}%
  \BibitemOpen
  \bibfield  {author} {\bibinfo {author} {\bibfnamefont {C.}~\bibnamefont
  {Ratti}},\ }\href {https://doi.org/https://doi.org/10.1088/1361-6633/aabb97}
  {\bibfield  {journal} {\bibinfo  {journal} {Reports on Progress in Physics}\
  }\textbf {\bibinfo {volume} {81}},\ \bibinfo {pages} {084301} (\bibinfo
  {year} {2018})}\BibitemShut {NoStop}%
\bibitem [{\citenamefont {Bernhard}\ \emph {et~al.}(2019)\citenamefont
  {Bernhard}, \citenamefont {Moreland},\ and\ \citenamefont
  {Bass}}]{bernhard2019bayesian}%
  \BibitemOpen
  \bibfield  {author} {\bibinfo {author} {\bibfnamefont {J.~E.}\ \bibnamefont
  {Bernhard}}, \bibinfo {author} {\bibfnamefont {J.~S.}\ \bibnamefont
  {Moreland}},\ and\ \bibinfo {author} {\bibfnamefont {S.~A.}\ \bibnamefont
  {Bass}},\ }\href {https://doi.org/https://doi.org/10.1038/s41567-019-0611-8}
  {\bibfield  {journal} {\bibinfo  {journal} {Nature Physics}\ }\textbf
  {\bibinfo {volume} {15}},\ \bibinfo {pages} {1113} (\bibinfo {year}
  {2019})}\BibitemShut {NoStop}%
\bibitem [{\citenamefont {Heinz}(2004)}]{heinz2004thermalization}%
  \BibitemOpen
  \bibfield  {author} {\bibinfo {author} {\bibfnamefont {U.}~\bibnamefont
  {Heinz}},\ }in\ \href {https://doi.org/https://doi.org/10.1063/1.1843595}
  {\emph {\bibinfo {booktitle} {AIP Conference Proceedings}}},\ Vol.\ \bibinfo
  {volume} {739}\ (\bibinfo {organization} {American Institute of Physics},\
  \bibinfo {year} {2004})\ pp.\ \bibinfo {pages} {163--180}\BibitemShut
  {NoStop}%
\bibitem [{\citenamefont {Martinez}\ and\ \citenamefont
  {Strickland}(2008)}]{martinez2008dilepton}%
  \BibitemOpen
  \bibfield  {author} {\bibinfo {author} {\bibfnamefont {M.}~\bibnamefont
  {Martinez}}\ and\ \bibinfo {author} {\bibfnamefont {M.}~\bibnamefont
  {Strickland}},\ }\href
  {https://doi.org/http://dx.doi.org/10.1088/0954-3899/35/10/104162} {\bibfield
   {journal} {\bibinfo  {journal} {Journal of Physics G: Nuclear and Particle
  Physics}\ }\textbf {\bibinfo {volume} {35}},\ \bibinfo {pages} {104162}
  (\bibinfo {year} {2008})}\BibitemShut {NoStop}%
\bibitem [{\citenamefont {Dubla}\ \emph {et~al.}(2018)\citenamefont {Dubla},
  \citenamefont {Masciocchi}, \citenamefont {Pawlowski}, \citenamefont
  {Schenke}, \citenamefont {Shen},\ and\ \citenamefont
  {Stachel}}]{dubla2018towards}%
  \BibitemOpen
  \bibfield  {author} {\bibinfo {author} {\bibfnamefont {A.}~\bibnamefont
  {Dubla}}, \bibinfo {author} {\bibfnamefont {S.}~\bibnamefont {Masciocchi}},
  \bibinfo {author} {\bibfnamefont {J.}~\bibnamefont {Pawlowski}}, \bibinfo
  {author} {\bibfnamefont {B.}~\bibnamefont {Schenke}}, \bibinfo {author}
  {\bibfnamefont {C.}~\bibnamefont {Shen}},\ and\ \bibinfo {author}
  {\bibfnamefont {J.}~\bibnamefont {Stachel}},\ }\href
  {https://doi.org/https://doi.org/10.1016/j.nuclphysa.2018.09.046} {\bibfield
  {journal} {\bibinfo  {journal} {Nuclear Physics A}\ }\textbf {\bibinfo
  {volume} {979}},\ \bibinfo {pages} {251} (\bibinfo {year}
  {2018})}\BibitemShut {NoStop}%
\bibitem [{\citenamefont {Bazavov}\ \emph {et~al.}(2019)\citenamefont
  {Bazavov}, \citenamefont {Ding}, \citenamefont {Hegde}, \citenamefont
  {Kaczmarek}, \citenamefont {Karsch}, \citenamefont {Karthik}, \citenamefont
  {Laermann}, \citenamefont {Lahiri}, \citenamefont {Larsen}, \citenamefont
  {Li} \emph {et~al.}}]{bazavov2019chiral}%
  \BibitemOpen
  \bibfield  {author} {\bibinfo {author} {\bibfnamefont {A.}~\bibnamefont
  {Bazavov}}, \bibinfo {author} {\bibfnamefont {H.-T.}\ \bibnamefont {Ding}},
  \bibinfo {author} {\bibfnamefont {P.}~\bibnamefont {Hegde}}, \bibinfo
  {author} {\bibfnamefont {O.}~\bibnamefont {Kaczmarek}}, \bibinfo {author}
  {\bibfnamefont {F.}~\bibnamefont {Karsch}}, \bibinfo {author} {\bibfnamefont
  {N.}~\bibnamefont {Karthik}}, \bibinfo {author} {\bibfnamefont
  {E.}~\bibnamefont {Laermann}}, \bibinfo {author} {\bibfnamefont
  {A.}~\bibnamefont {Lahiri}}, \bibinfo {author} {\bibfnamefont
  {R.}~\bibnamefont {Larsen}}, \bibinfo {author} {\bibfnamefont {S.-T.}\
  \bibnamefont {Li}}, \emph {et~al.},\ }\href
  {https://doi.org/https://doi.org/10.1016/j.physletb.2019.05.013} {\bibfield
  {journal} {\bibinfo  {journal} {Physics Letters B}\ }\textbf {\bibinfo
  {volume} {795}},\ \bibinfo {pages} {15} (\bibinfo {year} {2019})}\BibitemShut
  {NoStop}%
\bibitem [{\citenamefont {Steinbrecher}\ and\ \citenamefont
  {HotQCD-Collaboration}(2019)}]{steinbrecher2019qcd}%
  \BibitemOpen
  \bibfield  {author} {\bibinfo {author} {\bibfnamefont {P.}~\bibnamefont
  {Steinbrecher}}\ and\ \bibinfo {author} {\bibnamefont
  {HotQCD-Collaboration}},\ }\href
  {https://doi.org/https://doi.org/10.1016/j.nuclphysa.2018.08.025} {\bibfield
  {journal} {\bibinfo  {journal} {Nuclear Physics A}\ }\textbf {\bibinfo
  {volume} {982}},\ \bibinfo {pages} {847} (\bibinfo {year}
  {2019})}\BibitemShut {NoStop}%
\bibitem [{\citenamefont {D'Elia}(2019)}]{d2019high}%
  \BibitemOpen
  \bibfield  {author} {\bibinfo {author} {\bibfnamefont {M.}~\bibnamefont
  {D'Elia}},\ }\href
  {https://doi.org/https://doi.org/10.1016/j.nuclphysa.2018.10.042} {\bibfield
  {journal} {\bibinfo  {journal} {Nuclear Physics A}\ }\textbf {\bibinfo
  {volume} {982}},\ \bibinfo {pages} {99} (\bibinfo {year} {2019})}\BibitemShut
  {NoStop}%
\bibitem [{\citenamefont {Bass}\ \emph {et~al.}(1999)\citenamefont {Bass},
  \citenamefont {Gyulassy}, \citenamefont {Stoecker},\ and\ \citenamefont
  {Greiner}}]{bass1999signatures}%
  \BibitemOpen
  \bibfield  {author} {\bibinfo {author} {\bibfnamefont {S.~A.}\ \bibnamefont
  {Bass}}, \bibinfo {author} {\bibfnamefont {M.}~\bibnamefont {Gyulassy}},
  \bibinfo {author} {\bibfnamefont {H.}~\bibnamefont {Stoecker}},\ and\
  \bibinfo {author} {\bibfnamefont {W.}~\bibnamefont {Greiner}},\ }\href
  {https://doi.org/10.1088/0954-3899/25/3/013} {\bibfield  {journal} {\bibinfo
  {journal} {Journal of Physics G: Nuclear and Particle Physics}\ }\textbf
  {\bibinfo {volume} {25}},\ \bibinfo {pages} {R1} (\bibinfo {year}
  {1999})}\BibitemShut {NoStop}%
\bibitem [{\citenamefont {Matsui}\ and\ \citenamefont
  {Satz}(1986)}]{matsui1986}%
  \BibitemOpen
  \bibfield  {author} {\bibinfo {author} {\bibfnamefont {T.}~\bibnamefont
  {Matsui}}\ and\ \bibinfo {author} {\bibfnamefont {H.}~\bibnamefont {Satz}},\
  }\href
  {https://cds.cern.ch/record/169304/files/BNL-38344.pdf?subformat=pdfa&version=1}
  {\bibfield  {journal} {\bibinfo  {journal} {Physics Letters B}\ }\textbf
  {\bibinfo {volume} {178}},\ \bibinfo {pages} {416} (\bibinfo {year}
  {1986})}\BibitemShut {NoStop}%
\bibitem [{\citenamefont {Aaij}\ \emph {et~al.}(2014)\citenamefont {Aaij},
  \citenamefont {Adeva}, \citenamefont {Adinolfi}, \citenamefont {Adrover},
  \citenamefont {Affolder}, \citenamefont {Ajaltouni}, \citenamefont
  {Albrecht}, \citenamefont {Alessio}, \citenamefont {Alexander}, \citenamefont
  {Ali} \emph {et~al.}}]{aaij2014CMN}%
  \BibitemOpen
  \bibfield  {author} {\bibinfo {author} {\bibfnamefont {R.}~\bibnamefont
  {Aaij}}, \bibinfo {author} {\bibfnamefont {B.}~\bibnamefont {Adeva}},
  \bibinfo {author} {\bibfnamefont {M.}~\bibnamefont {Adinolfi}}, \bibinfo
  {author} {\bibfnamefont {C.}~\bibnamefont {Adrover}}, \bibinfo {author}
  {\bibfnamefont {A.}~\bibnamefont {Affolder}}, \bibinfo {author}
  {\bibfnamefont {Z.}~\bibnamefont {Ajaltouni}}, \bibinfo {author}
  {\bibfnamefont {J.}~\bibnamefont {Albrecht}}, \bibinfo {author}
  {\bibfnamefont {F.}~\bibnamefont {Alessio}}, \bibinfo {author} {\bibfnamefont
  {M.}~\bibnamefont {Alexander}}, \bibinfo {author} {\bibfnamefont
  {S.}~\bibnamefont {Ali}}, \emph {et~al.},\ }\href
  {https://doi.org/http://dx.doi.org/10.1007/JHEP02(2014)072} {\bibfield
  {journal} {\bibinfo  {journal} {Journal of High Energy Physics}\ }\textbf
  {\bibinfo {volume} {2014}},\ \bibinfo {pages} {72} (\bibinfo {year}
  {2014})}\BibitemShut {NoStop}%
\bibitem [{\citenamefont {Schukraft}(2017)}]{schukraft2017qm2017}%
  \BibitemOpen
  \bibfield  {author} {\bibinfo {author} {\bibfnamefont {J.}~\bibnamefont
  {Schukraft}},\ }\href
  {https://doi.org/https://doi.org/10.1016/j.nuclphysa.2017.05.036} {\bibfield
  {journal} {\bibinfo  {journal} {Nuclear Physics A}\ }\textbf {\bibinfo
  {volume} {967}},\ \bibinfo {pages} {1} (\bibinfo {year} {2017})}\BibitemShut
  {NoStop}%
\bibitem [{\citenamefont {Sharma}\ and\ \citenamefont
  {Vitev}(2013)}]{vitev2013}%
  \BibitemOpen
  \bibfield  {author} {\bibinfo {author} {\bibfnamefont {R.}~\bibnamefont
  {Sharma}}\ and\ \bibinfo {author} {\bibfnamefont {I.}~\bibnamefont {Vitev}},\
  }\href {https://doi.org/http://dx.doi.org/10.1103/PhysRevC.87.044905}
  {\bibfield  {journal} {\bibinfo  {journal} {Physical Review C}\ }\textbf
  {\bibinfo {volume} {87}},\ \bibinfo {pages} {044905} (\bibinfo {year}
  {2013})}\BibitemShut {NoStop}%
\bibitem [{\citenamefont {Du}\ \emph {et~al.}(2017)\citenamefont {Du},
  \citenamefont {He},\ and\ \citenamefont {Rapp}}]{Rapp2017}%
  \BibitemOpen
  \bibfield  {author} {\bibinfo {author} {\bibfnamefont {X.}~\bibnamefont
  {Du}}, \bibinfo {author} {\bibfnamefont {M.}~\bibnamefont {He}},\ and\
  \bibinfo {author} {\bibfnamefont {R.}~\bibnamefont {Rapp}},\ }\href
  {https://doi.org/https://doi.org/10.1103/PhysRevC.96.054901} {\bibfield
  {journal} {\bibinfo  {journal} {Physical Review C}\ }\textbf {\bibinfo
  {volume} {96}},\ \bibinfo {pages} {054901} (\bibinfo {year}
  {2017})}\BibitemShut {NoStop}%
\bibitem [{\citenamefont {Krouppa}\ \emph {et~al.}(2018)\citenamefont
  {Krouppa}, \citenamefont {Rothkopf},\ and\ \citenamefont
  {Strickland}}]{strickland2018}%
  \BibitemOpen
  \bibfield  {author} {\bibinfo {author} {\bibfnamefont {B.}~\bibnamefont
  {Krouppa}}, \bibinfo {author} {\bibfnamefont {A.}~\bibnamefont {Rothkopf}},\
  and\ \bibinfo {author} {\bibfnamefont {M.}~\bibnamefont {Strickland}},\
  }\href {https://doi.org/https://doi.org/10.1103/PhysRevD.97.016017}
  {\bibfield  {journal} {\bibinfo  {journal} {Physical Review D}\ }\textbf
  {\bibinfo {volume} {97}},\ \bibinfo {pages} {016017} (\bibinfo {year}
  {2018})}\BibitemShut {NoStop}%
\bibitem [{\citenamefont {Digal}\ \emph {et~al.}(2001)\citenamefont {Digal},
  \citenamefont {Petreczky},\ and\ \citenamefont {Satz}}]{digal2001quarkonium}%
  \BibitemOpen
  \bibfield  {author} {\bibinfo {author} {\bibfnamefont {S.}~\bibnamefont
  {Digal}}, \bibinfo {author} {\bibfnamefont {P.}~\bibnamefont {Petreczky}},\
  and\ \bibinfo {author} {\bibfnamefont {H.}~\bibnamefont {Satz}},\ }\href
  {https://doi.org/10.1103/PhysRevD.64.094015} {\bibfield  {journal} {\bibinfo
  {journal} {Physical Review D}\ }\textbf {\bibinfo {volume} {64}},\ \bibinfo
  {pages} {094015} (\bibinfo {year} {2001})}\BibitemShut {NoStop}%
\bibitem [{\citenamefont {Singh}\ \emph {et~al.}(2019)\citenamefont {Singh},
  \citenamefont {Ganesh},\ and\ \citenamefont {Mishra}}]{singh2019}%
  \BibitemOpen
  \bibfield  {author} {\bibinfo {author} {\bibfnamefont {C.~R.}\ \bibnamefont
  {Singh}}, \bibinfo {author} {\bibfnamefont {S.}~\bibnamefont {Ganesh}},\ and\
  \bibinfo {author} {\bibfnamefont {M.}~\bibnamefont {Mishra}},\ }\href
  {https://doi.org/https://doi.org/10.1140/epjc/s10052-019-6646-2} {\bibfield
  {journal} {\bibinfo  {journal} {The European Physical Journal C}\ }\textbf
  {\bibinfo {volume} {79}},\ \bibinfo {pages} {147} (\bibinfo {year}
  {2019})}\BibitemShut {NoStop}%
\bibitem [{\citenamefont {Ganesh}\ and\ \citenamefont
  {Mishra}(2013)}]{bottomganesh2013}%
  \BibitemOpen
  \bibfield  {author} {\bibinfo {author} {\bibfnamefont {S.}~\bibnamefont
  {Ganesh}}\ and\ \bibinfo {author} {\bibfnamefont {M.}~\bibnamefont
  {Mishra}},\ }\href {https://doi.org/10.1103/PhysRevC.88.044908} {\bibfield
  {journal} {\bibinfo  {journal} {Phys. Rev. C}\ }\textbf {\bibinfo {volume}
  {88}},\ \bibinfo {pages} {044908} (\bibinfo {year} {2013})}\BibitemShut
  {NoStop}%
\bibitem [{\citenamefont {Srivastava}\ \emph {et~al.}(2013)\citenamefont
  {Srivastava}, \citenamefont {Mishra},\ and\ \citenamefont
  {Singh}}]{qpmcolorscreening2013}%
  \BibitemOpen
  \bibfield  {author} {\bibinfo {author} {\bibfnamefont {P.~K.}\ \bibnamefont
  {Srivastava}}, \bibinfo {author} {\bibfnamefont {M.}~\bibnamefont {Mishra}},\
  and\ \bibinfo {author} {\bibfnamefont {C.~P.}\ \bibnamefont {Singh}},\ }\href
  {https://doi.org/10.1103/PhysRevC.87.034903} {\bibfield  {journal} {\bibinfo
  {journal} {Phys. Rev. C}\ }\textbf {\bibinfo {volume} {87}},\ \bibinfo
  {pages} {034903} (\bibinfo {year} {2013})}\BibitemShut {NoStop}%
\bibitem [{\citenamefont {Mishra}\ \emph {et~al.}(2007)\citenamefont {Mishra},
  \citenamefont {Singh}, \citenamefont {Menon},\ and\ \citenamefont
  {Dubey}}]{MishraCS2007}%
  \BibitemOpen
  \bibfield  {author} {\bibinfo {author} {\bibfnamefont {M.}~\bibnamefont
  {Mishra}}, \bibinfo {author} {\bibfnamefont {C.}~\bibnamefont {Singh}},
  \bibinfo {author} {\bibfnamefont {V.}~\bibnamefont {Menon}},\ and\ \bibinfo
  {author} {\bibfnamefont {R.~K.}\ \bibnamefont {Dubey}},\ }\href
  {https://doi.org/https://doi.org/10.1016/j.physletb.2007.09.043} {\bibfield
  {journal} {\bibinfo  {journal} {Physics Letters B}\ }\textbf {\bibinfo
  {volume} {656}},\ \bibinfo {pages} {45 } (\bibinfo {year}
  {2007})}\BibitemShut {NoStop}%
\bibitem [{\citenamefont {Krouppa}\ \emph {et~al.}(2019)\citenamefont
  {Krouppa}, \citenamefont {Rothkopf},\ and\ \citenamefont
  {Strickland}}]{krouppa2019bottomonium}%
  \BibitemOpen
  \bibfield  {author} {\bibinfo {author} {\bibfnamefont {B.}~\bibnamefont
  {Krouppa}}, \bibinfo {author} {\bibfnamefont {A.}~\bibnamefont {Rothkopf}},\
  and\ \bibinfo {author} {\bibfnamefont {M.}~\bibnamefont {Strickland}},\
  }\href {https://doi.org/https://doi.org/10.1016/j.nuclphysa.2018.09.034}
  {\bibfield  {journal} {\bibinfo  {journal} {Nuclear Physics A}\ }\textbf
  {\bibinfo {volume} {982}},\ \bibinfo {pages} {727} (\bibinfo {year}
  {2019})}\BibitemShut {NoStop}%
\bibitem [{\citenamefont {McDonald}\ \emph {et~al.}(2017)\citenamefont
  {McDonald}, \citenamefont {Shen}, \citenamefont {Fillion-Gourdeau},
  \citenamefont {Jeon},\ and\ \citenamefont {Gale}}]{mcdonald2017hydrodynamic}%
  \BibitemOpen
  \bibfield  {author} {\bibinfo {author} {\bibfnamefont {S.}~\bibnamefont
  {McDonald}}, \bibinfo {author} {\bibfnamefont {C.}~\bibnamefont {Shen}},
  \bibinfo {author} {\bibfnamefont {F.}~\bibnamefont {Fillion-Gourdeau}},
  \bibinfo {author} {\bibfnamefont {S.}~\bibnamefont {Jeon}},\ and\ \bibinfo
  {author} {\bibfnamefont {C.}~\bibnamefont {Gale}},\ }\href
  {https://doi.org/10.1103/PhysRevC.93.034912} {\bibfield  {journal} {\bibinfo
  {journal} {Physical Review C}\ }\textbf {\bibinfo {volume} {95}},\ \bibinfo
  {pages} {064913} (\bibinfo {year} {2017})}\BibitemShut {NoStop}%
\bibitem [{\citenamefont {Alqahtani}\ \emph {et~al.}(2017)\citenamefont
  {Alqahtani}, \citenamefont {Nopoush}, \citenamefont {Ryblewski},\ and\
  \citenamefont {Strickland}}]{alqahtani20173}%
  \BibitemOpen
  \bibfield  {author} {\bibinfo {author} {\bibfnamefont {M.}~\bibnamefont
  {Alqahtani}}, \bibinfo {author} {\bibfnamefont {M.}~\bibnamefont {Nopoush}},
  \bibinfo {author} {\bibfnamefont {R.}~\bibnamefont {Ryblewski}},\ and\
  \bibinfo {author} {\bibfnamefont {M.}~\bibnamefont {Strickland}},\ }\href
  {https://doi.org/10.1103/PhysRevLett.119.042301} {\bibfield  {journal}
  {\bibinfo  {journal} {Physical review letters}\ }\textbf {\bibinfo {volume}
  {119}},\ \bibinfo {pages} {042301} (\bibinfo {year} {2017})}\BibitemShut
  {NoStop}%
\bibitem [{\citenamefont {Habich}\ \emph {et~al.}(2015)\citenamefont {Habich},
  \citenamefont {Nagle},\ and\ \citenamefont
  {Romatschke}}]{habich2015particle}%
  \BibitemOpen
  \bibfield  {author} {\bibinfo {author} {\bibfnamefont {M.}~\bibnamefont
  {Habich}}, \bibinfo {author} {\bibfnamefont {J.}~\bibnamefont {Nagle}},\ and\
  \bibinfo {author} {\bibfnamefont {P.}~\bibnamefont {Romatschke}},\ }\href
  {https://doi.org/10.1140/epjc/s10052-014-3206-7} {\bibfield  {journal}
  {\bibinfo  {journal} {The European Physical Journal C}\ }\textbf {\bibinfo
  {volume} {75}},\ \bibinfo {pages} {15} (\bibinfo {year} {2015})}\BibitemShut
  {NoStop}%
\bibitem [{\citenamefont {Bjorken}(1983)}]{bjorken1983highly}%
  \BibitemOpen
  \bibfield  {author} {\bibinfo {author} {\bibfnamefont {J.~D.}\ \bibnamefont
  {Bjorken}},\ }\href {https://doi.org/https://doi.org/10.1103/PhysRevD.27.140}
  {\bibfield  {journal} {\bibinfo  {journal} {Physical review D}\ }\textbf
  {\bibinfo {volume} {27}},\ \bibinfo {pages} {140} (\bibinfo {year}
  {1983})}\BibitemShut {NoStop}%
\bibitem [{\citenamefont {Del~Zanna}\ \emph {et~al.}(2013)\citenamefont
  {Del~Zanna}, \citenamefont {Chandra}, \citenamefont {Inghirami},
  \citenamefont {Rolando}, \citenamefont {Beraudo}, \citenamefont {De~Pace},
  \citenamefont {Pagliara}, \citenamefont {Drago},\ and\ \citenamefont
  {Becattini}}]{del2013relativistic}%
  \BibitemOpen
  \bibfield  {author} {\bibinfo {author} {\bibfnamefont {L.}~\bibnamefont
  {Del~Zanna}}, \bibinfo {author} {\bibfnamefont {V.}~\bibnamefont {Chandra}},
  \bibinfo {author} {\bibfnamefont {G.}~\bibnamefont {Inghirami}}, \bibinfo
  {author} {\bibfnamefont {V.}~\bibnamefont {Rolando}}, \bibinfo {author}
  {\bibfnamefont {A.}~\bibnamefont {Beraudo}}, \bibinfo {author} {\bibfnamefont
  {A.}~\bibnamefont {De~Pace}}, \bibinfo {author} {\bibfnamefont
  {G.}~\bibnamefont {Pagliara}}, \bibinfo {author} {\bibfnamefont
  {A.}~\bibnamefont {Drago}},\ and\ \bibinfo {author} {\bibfnamefont
  {F.}~\bibnamefont {Becattini}},\ }\href
  {https://doi.org/10.1140/epjc/s10052-013-2524-5} {\bibfield  {journal}
  {\bibinfo  {journal} {The European Physical Journal C}\ }\textbf {\bibinfo
  {volume} {73}},\ \bibinfo {pages} {2524} (\bibinfo {year}
  {2013})}\BibitemShut {NoStop}%
\bibitem [{\citenamefont {Borsanyi}\ \emph {et~al.}(2010)\citenamefont
  {Borsanyi}, \citenamefont {Endr{\H{o}}di}, \citenamefont {Fodor},
  \citenamefont {Jakovac}, \citenamefont {Katz}, \citenamefont {Krieg},
  \citenamefont {Ratti},\ and\ \citenamefont {Szabo}}]{WBEoS2010}%
  \BibitemOpen
  \bibfield  {author} {\bibinfo {author} {\bibfnamefont {S.}~\bibnamefont
  {Borsanyi}}, \bibinfo {author} {\bibfnamefont {G.}~\bibnamefont
  {Endr{\H{o}}di}}, \bibinfo {author} {\bibfnamefont {Z.}~\bibnamefont
  {Fodor}}, \bibinfo {author} {\bibfnamefont {A.}~\bibnamefont {Jakovac}},
  \bibinfo {author} {\bibfnamefont {S.~D.}\ \bibnamefont {Katz}}, \bibinfo
  {author} {\bibfnamefont {S.}~\bibnamefont {Krieg}}, \bibinfo {author}
  {\bibfnamefont {C.}~\bibnamefont {Ratti}},\ and\ \bibinfo {author}
  {\bibfnamefont {K.~K.}\ \bibnamefont {Szabo}},\ }\href
  {https://doi.org/10.1007/JHEP11(2010)077} {\bibfield  {journal} {\bibinfo
  {journal} {Journal of High Energy Physics}\ }\textbf {\bibinfo {volume}
  {2010}},\ \bibinfo {pages} {77} (\bibinfo {year} {2010})}\BibitemShut
  {NoStop}%
\bibitem [{\citenamefont {Srivastava}\ \emph {et~al.}(2010)\citenamefont
  {Srivastava}, \citenamefont {Tiwari},\ and\ \citenamefont {Singh}}]{qpm2010}%
  \BibitemOpen
  \bibfield  {author} {\bibinfo {author} {\bibfnamefont {P.~K.}\ \bibnamefont
  {Srivastava}}, \bibinfo {author} {\bibfnamefont {S.~K.}\ \bibnamefont
  {Tiwari}},\ and\ \bibinfo {author} {\bibfnamefont {C.~P.}\ \bibnamefont
  {Singh}},\ }\href {https://doi.org/10.1103/PhysRevD.82.014023} {\bibfield
  {journal} {\bibinfo  {journal} {Phys. Rev. D}\ }\textbf {\bibinfo {volume}
  {82}},\ \bibinfo {pages} {014023} (\bibinfo {year} {2010})}\BibitemShut
  {NoStop}%
\bibitem [{\citenamefont {Chatterjee}\ \emph {et~al.}(2010)\citenamefont
  {Chatterjee}, \citenamefont {Godbole},\ and\ \citenamefont
  {Gupta}}]{hrg2010}%
  \BibitemOpen
  \bibfield  {author} {\bibinfo {author} {\bibfnamefont {S.}~\bibnamefont
  {Chatterjee}}, \bibinfo {author} {\bibfnamefont {R.}~\bibnamefont
  {Godbole}},\ and\ \bibinfo {author} {\bibfnamefont {S.}~\bibnamefont
  {Gupta}},\ }\href {https://doi.org/10.1103/PhysRevC.81.044907} {\bibfield
  {journal} {\bibinfo  {journal} {Physical Review C}\ }\textbf {\bibinfo
  {volume} {81}},\ \bibinfo {pages} {044907} (\bibinfo {year}
  {2010})}\BibitemShut {NoStop}%
\bibitem [{\citenamefont {Ding}(2020)}]{ding2020new}%
  \BibitemOpen
  \bibfield  {author} {\bibinfo {author} {\bibfnamefont {H.-T.}\ \bibnamefont
  {Ding}},\ }\bibfield  {journal} {\bibinfo  {journal} {arXiv preprint
  arXiv:2002.11957}\ }\href {https://doi.org/https://arxiv.org/abs/2002.11957}
  {https://arxiv.org/abs/2002.11957} (\bibinfo {year} {2020})\BibitemShut
  {NoStop}%
\bibitem [{\citenamefont {Loizides}\ \emph {et~al.}(2018)\citenamefont
  {Loizides}, \citenamefont {Kamin},\ and\ \citenamefont
  {d'Enterria}}]{loizides2018improved}%
  \BibitemOpen
  \bibfield  {author} {\bibinfo {author} {\bibfnamefont {C.}~\bibnamefont
  {Loizides}}, \bibinfo {author} {\bibfnamefont {J.}~\bibnamefont {Kamin}},\
  and\ \bibinfo {author} {\bibfnamefont {D.}~\bibnamefont {d'Enterria}},\
  }\href {https://doi.org/https://doi.org/10.1103/PhysRevC.97.054910}
  {\bibfield  {journal} {\bibinfo  {journal} {Physical Review C}\ }\textbf
  {\bibinfo {volume} {97}},\ \bibinfo {pages} {054910} (\bibinfo {year}
  {2018})}\BibitemShut {NoStop}%
\bibitem [{\citenamefont {et. al.}(2013)}]{2p76rap_spectra2013}%
  \BibitemOpen
  \bibfield  {author} {\bibinfo {author} {\bibfnamefont {E.~A.}\ \bibnamefont
  {et. al.}},\ }\href
  {https://doi.org/https://doi.org/10.1016/j.physletb.2013.09.022} {\bibfield
  {journal} {\bibinfo  {journal} {Physics Letters B}\ }\textbf {\bibinfo
  {volume} {726}},\ \bibinfo {pages} {610 } (\bibinfo {year}
  {2013})}\BibitemShut {NoStop}%
\bibitem [{\citenamefont {et. al.}(2017)}]{5p02rap_spectra2017}%
  \BibitemOpen
  \bibfield  {author} {\bibinfo {author} {\bibfnamefont {J.~A.}\ \bibnamefont
  {et. al.}},\ }\href
  {https://doi.org/https://doi.org/10.1016/j.physletb.2017.07.017} {\bibfield
  {journal} {\bibinfo  {journal} {Physics Letters B}\ }\textbf {\bibinfo
  {volume} {772}},\ \bibinfo {pages} {567 } (\bibinfo {year}
  {2017})}\BibitemShut {NoStop}%
\bibitem [{\citenamefont {Busza}\ \emph {et~al.}(2018)\citenamefont {Busza},
  \citenamefont {Rajagopal},\ and\ \citenamefont {Van
  Der~Schee}}]{busza2018heavy}%
  \BibitemOpen
  \bibfield  {author} {\bibinfo {author} {\bibfnamefont {W.}~\bibnamefont
  {Busza}}, \bibinfo {author} {\bibfnamefont {K.}~\bibnamefont {Rajagopal}},\
  and\ \bibinfo {author} {\bibfnamefont {W.}~\bibnamefont {Van Der~Schee}},\
  }\href {https://doi.org/https://doi.org/10.1146/annurev-nucl-101917-020852}
  {\bibfield  {journal} {\bibinfo  {journal} {Annual Review of Nuclear and
  Particle Science}\ }\textbf {\bibinfo {volume} {68}},\ \bibinfo {pages} {339}
  (\bibinfo {year} {2018})}\BibitemShut {NoStop}%
\bibitem [{\citenamefont {Song}\ and\ \citenamefont
  {Heinz}(2010)}]{song2010interplay}%
  \BibitemOpen
  \bibfield  {author} {\bibinfo {author} {\bibfnamefont {H.}~\bibnamefont
  {Song}}\ and\ \bibinfo {author} {\bibfnamefont {U.}~\bibnamefont {Heinz}},\
  }\href {https://doi.org/10.1103/PhysRevC.81.024905} {\bibfield  {journal}
  {\bibinfo  {journal} {Physical Review C}\ }\textbf {\bibinfo {volume} {81}},\
  \bibinfo {pages} {024905} (\bibinfo {year} {2010})}\BibitemShut {NoStop}%
\bibitem [{\citenamefont {Bhaduri}\ \emph {et~al.}(2019)\citenamefont
  {Bhaduri}, \citenamefont {Borghini}, \citenamefont {Jaiswal},\ and\
  \citenamefont {Strickland}}]{bhaduri2019anisotropic}%
  \BibitemOpen
  \bibfield  {author} {\bibinfo {author} {\bibfnamefont {P.~P.}\ \bibnamefont
  {Bhaduri}}, \bibinfo {author} {\bibfnamefont {N.}~\bibnamefont {Borghini}},
  \bibinfo {author} {\bibfnamefont {A.}~\bibnamefont {Jaiswal}},\ and\ \bibinfo
  {author} {\bibfnamefont {M.}~\bibnamefont {Strickland}},\ }\href
  {https://doi.org/https://doi.org/10.1103/PhysRevC.100.051901} {\bibfield
  {journal} {\bibinfo  {journal} {Physical Review C}\ }\textbf {\bibinfo
  {volume} {100}},\ \bibinfo {pages} {051901(R)} (\bibinfo {year}
  {2019})}\BibitemShut {NoStop}%
\bibitem [{\citenamefont {Karsch}(2002)}]{KarschLQCD2002}%
  \BibitemOpen
  \bibfield  {author} {\bibinfo {author} {\bibfnamefont {F.}~\bibnamefont
  {Karsch}},\ }\href {https://doi.org/https://doi.org/10.1007/3-540-45792-5_6}
  {\emph {\bibinfo {title} {Lattice QCD at high temperature and density, in
  Lectures on Quark Matter, Lecture Notes in Physics}}},\ Vol.\ \bibinfo
  {volume} {583}\ (\bibinfo  {publisher} {Springer, Berlin, Heidelberg},\
  \bibinfo {year} {2002})\ p.\ \bibinfo {pages} {209}\BibitemShut {NoStop}%
\bibitem [{\citenamefont {Adler}\ \emph {et~al.}(2005)\citenamefont {Adler},
  \citenamefont {Afanasiev}, \citenamefont {Aidala}, \citenamefont {Ajitanand},
  \citenamefont {Akiba}, \citenamefont {Alexander}, \citenamefont {Amirikas},
  \citenamefont {Aphecetche}, \citenamefont {Aronson}, \citenamefont {Averbeck}
  \emph {et~al.}}]{adler2005systematic}%
  \BibitemOpen
  \bibfield  {author} {\bibinfo {author} {\bibfnamefont {S.~S.}\ \bibnamefont
  {Adler}}, \bibinfo {author} {\bibfnamefont {S.}~\bibnamefont {Afanasiev}},
  \bibinfo {author} {\bibfnamefont {C.}~\bibnamefont {Aidala}}, \bibinfo
  {author} {\bibfnamefont {N.}~\bibnamefont {Ajitanand}}, \bibinfo {author}
  {\bibfnamefont {Y.}~\bibnamefont {Akiba}}, \bibinfo {author} {\bibfnamefont
  {J.}~\bibnamefont {Alexander}}, \bibinfo {author} {\bibfnamefont
  {R.}~\bibnamefont {Amirikas}}, \bibinfo {author} {\bibfnamefont
  {L.}~\bibnamefont {Aphecetche}}, \bibinfo {author} {\bibfnamefont
  {S.}~\bibnamefont {Aronson}}, \bibinfo {author} {\bibfnamefont
  {R.}~\bibnamefont {Averbeck}}, \emph {et~al.},\ }\href
  {https://doi.org/https://doi.org/10.1103/PhysRevC.71.034908} {\bibfield
  {journal} {\bibinfo  {journal} {Physical Review C}\ }\textbf {\bibinfo
  {volume} {71}},\ \bibinfo {pages} {034908} (\bibinfo {year}
  {2005})}\BibitemShut {NoStop}%
\bibitem [{\citenamefont {Acharya}(2020)}]{5p02pT_spectra2020}%
  \BibitemOpen
  \bibfield  {author} {\bibinfo {author} {\bibfnamefont {S.~e.~a.}\
  \bibnamefont {Acharya}} (\bibinfo {collaboration} {ALICE Collaboration}),\
  }\href {https://doi.org/10.1103/PhysRevC.101.044907} {\bibfield  {journal}
  {\bibinfo  {journal} {Phys. Rev. C}\ }\textbf {\bibinfo {volume} {101}},\
  \bibinfo {pages} {044907} (\bibinfo {year} {2020})}\BibitemShut {NoStop}%
\bibitem [{\citenamefont {Adam}(2016)}]{2p76pT_spectra2017}%
  \BibitemOpen
  \bibfield  {author} {\bibinfo {author} {\bibfnamefont {J.~e.~a.}\
  \bibnamefont {Adam}} (\bibinfo {collaboration} {ALICE Collaboration}),\
  }\href {https://doi.org/10.1103/PhysRevC.93.034913} {\bibfield  {journal}
  {\bibinfo  {journal} {Phys. Rev. C}\ }\textbf {\bibinfo {volume} {93}},\
  \bibinfo {pages} {034913} (\bibinfo {year} {2016})}\BibitemShut {NoStop}%
\bibitem [{\citenamefont {Nendzig}\ and\ \citenamefont
  {Wolschin}(2013)}]{Nendzig2013}%
  \BibitemOpen
  \bibfield  {author} {\bibinfo {author} {\bibfnamefont {F.}~\bibnamefont
  {Nendzig}}\ and\ \bibinfo {author} {\bibfnamefont {G.}~\bibnamefont
  {Wolschin}},\ }\href {https://doi.org/10.1103/PhysRevC.87.024911} {\bibfield
  {journal} {\bibinfo  {journal} {Phys. Rev. C}\ }\textbf {\bibinfo {volume}
  {87}},\ \bibinfo {pages} {024911} (\bibinfo {year} {2013})}\BibitemShut
  {NoStop}%
\bibitem [{\citenamefont {Brambilla}\ \emph {et~al.}(2011)\citenamefont
  {Brambilla}, \citenamefont {Escobedo}, \citenamefont {Ghiglieri},\ and\
  \citenamefont {Vairo}}]{brambilla2011_gluo}%
  \BibitemOpen
  \bibfield  {author} {\bibinfo {author} {\bibfnamefont {N.}~\bibnamefont
  {Brambilla}}, \bibinfo {author} {\bibfnamefont {M.~{\'A}.}\ \bibnamefont
  {Escobedo}}, \bibinfo {author} {\bibfnamefont {J.}~\bibnamefont
  {Ghiglieri}},\ and\ \bibinfo {author} {\bibfnamefont {A.}~\bibnamefont
  {Vairo}},\ }\href {https://doi.org/http://dx.doi.org/10.1007/JHEP12(2011)116}
  {\bibfield  {journal} {\bibinfo  {journal} {Journal of High Energy Physics}\
  }\textbf {\bibinfo {volume} {2011}},\ \bibinfo {pages} {1} (\bibinfo {year}
  {2011})}\BibitemShut {NoStop}%
\bibitem [{\citenamefont {Laine}\ \emph {et~al.}(2007)\citenamefont {Laine},
  \citenamefont {Philipsen}, \citenamefont {Tassler},\ and\ \citenamefont
  {Romatschke}}]{Laine2007}%
  \BibitemOpen
  \bibfield  {author} {\bibinfo {author} {\bibfnamefont {M.}~\bibnamefont
  {Laine}}, \bibinfo {author} {\bibfnamefont {O.}~\bibnamefont {Philipsen}},
  \bibinfo {author} {\bibfnamefont {M.}~\bibnamefont {Tassler}},\ and\ \bibinfo
  {author} {\bibfnamefont {P.}~\bibnamefont {Romatschke}},\ }\href
  {https://doi.org/10.1088/1126-6708/2007/03/054} {\bibfield  {journal}
  {\bibinfo  {journal} {Journal of High Energy Physics}\ }\textbf {\bibinfo
  {volume} {2007}},\ \bibinfo {pages} {054} (\bibinfo {year}
  {2007})}\BibitemShut {NoStop}%
\bibitem [{\citenamefont {Brambilla}\ \emph {et~al.}(2013)\citenamefont
  {Brambilla}, \citenamefont {Escobedo}, \citenamefont {Ghiglieri},\ and\
  \citenamefont {Vairo}}]{brambilla2013_damping}%
  \BibitemOpen
  \bibfield  {author} {\bibinfo {author} {\bibfnamefont {N.}~\bibnamefont
  {Brambilla}}, \bibinfo {author} {\bibfnamefont {M.~{\'A}.}\ \bibnamefont
  {Escobedo}}, \bibinfo {author} {\bibfnamefont {J.}~\bibnamefont
  {Ghiglieri}},\ and\ \bibinfo {author} {\bibfnamefont {A.}~\bibnamefont
  {Vairo}},\ }\href {https://doi.org/http://dx.doi.org/10.1007/JHEP05(2013)130}
  {\bibfield  {journal} {\bibinfo  {journal} {Journal of High Energy Physics}\
  }\textbf {\bibinfo {volume} {2013}},\ \bibinfo {pages} {130} (\bibinfo {year}
  {2013})}\BibitemShut {NoStop}%
\bibitem [{\citenamefont {Strickland}\ and\ \citenamefont
  {Bazow}(2011)}]{strickland2011thermal}%
  \BibitemOpen
  \bibfield  {author} {\bibinfo {author} {\bibfnamefont {M.}~\bibnamefont
  {Strickland}}\ and\ \bibinfo {author} {\bibfnamefont {D.}~\bibnamefont
  {Bazow}},\ }\bibfield  {journal} {\bibinfo  {journal} {arXiv preprint
  arXiv:1112.2761}\ }\href {https://doi.org/10.1016/j.nuclphysa.2012.02.003}
  {10.1016/j.nuclphysa.2012.02.003} (\bibinfo {year} {2011})\BibitemShut
  {NoStop}%
\bibitem [{\citenamefont {Loizides}\ \emph {et~al.}(2019)\citenamefont
  {Loizides}, \citenamefont {Kamin},\ and\ \citenamefont
  {d'Enterria}}]{loizides2019erratum}%
  \BibitemOpen
  \bibfield  {author} {\bibinfo {author} {\bibfnamefont {C.}~\bibnamefont
  {Loizides}}, \bibinfo {author} {\bibfnamefont {J.}~\bibnamefont {Kamin}},\
  and\ \bibinfo {author} {\bibfnamefont {D.}~\bibnamefont {d'Enterria}},\
  }\href {https://doi.org/https://doi.org/10.1103/PhysRevC.99.019901}
  {\bibfield  {journal} {\bibinfo  {journal} {Physical Review C}\ }\textbf
  {\bibinfo {volume} {99}},\ \bibinfo {pages} {019901(E)} (\bibinfo {year}
  {2019})}\BibitemShut {NoStop}%
\bibitem [{\citenamefont {Cheng}\ \emph {et~al.}(2018)\citenamefont {Cheng},
  \citenamefont {Meng}, \citenamefont {Xia}, \citenamefont {Ping},\ and\
  \citenamefont {Zong}}]{DissoTemp2018}%
  \BibitemOpen
  \bibfield  {author} {\bibinfo {author} {\bibfnamefont {P.}~\bibnamefont
  {Cheng}}, \bibinfo {author} {\bibfnamefont {Q.}~\bibnamefont {Meng}},
  \bibinfo {author} {\bibfnamefont {Y.}~\bibnamefont {Xia}}, \bibinfo {author}
  {\bibfnamefont {J.}~\bibnamefont {Ping}},\ and\ \bibinfo {author}
  {\bibfnamefont {H.}~\bibnamefont {Zong}},\ }\href
  {https://doi.org/10.1103/PhysRevD.98.116010} {\bibfield  {journal} {\bibinfo
  {journal} {Phys. Rev. D}\ }\textbf {\bibinfo {volume} {98}},\ \bibinfo
  {pages} {116010} (\bibinfo {year} {2018})}\BibitemShut {NoStop}%
\bibitem [{\citenamefont {Vogt}(2010)}]{vogt2010cold}%
  \BibitemOpen
  \bibfield  {author} {\bibinfo {author} {\bibfnamefont {R.}~\bibnamefont
  {Vogt}},\ }\href
  {https://doi.org/http://dx.doi.org/10.1103/PhysRevC.81.044903} {\bibfield
  {journal} {\bibinfo  {journal} {Physical Review C}\ }\textbf {\bibinfo
  {volume} {81}},\ \bibinfo {pages} {044903} (\bibinfo {year}
  {2010})}\BibitemShut {NoStop}%
\bibitem [{\citenamefont {Eskola}\ \emph {et~al.}(2009)\citenamefont {Eskola},
  \citenamefont {Paukkunen},\ and\ \citenamefont {Salgado}}]{eskola2009eps09}%
  \BibitemOpen
  \bibfield  {author} {\bibinfo {author} {\bibfnamefont {K.}~\bibnamefont
  {Eskola}}, \bibinfo {author} {\bibfnamefont {H.}~\bibnamefont {Paukkunen}},\
  and\ \bibinfo {author} {\bibfnamefont {C.}~\bibnamefont {Salgado}},\ }\href
  {https://doi.org/https://doi:10.1088/1126-6708/2009/04/065} {\bibfield
  {journal} {\bibinfo  {journal} {Journal of High Energy Physics}\ }\textbf
  {\bibinfo {volume} {2009}},\ \bibinfo {pages} {065} (\bibinfo {year}
  {2009})}\BibitemShut {NoStop}%
\bibitem [{\citenamefont {Eskola}\ \emph {et~al.}(2017)\citenamefont {Eskola},
  \citenamefont {Paakkinen}, \citenamefont {Paukkunen},\ and\ \citenamefont
  {Salgado}}]{eskola2017epps16}%
  \BibitemOpen
  \bibfield  {author} {\bibinfo {author} {\bibfnamefont {K.~J.}\ \bibnamefont
  {Eskola}}, \bibinfo {author} {\bibfnamefont {P.}~\bibnamefont {Paakkinen}},
  \bibinfo {author} {\bibfnamefont {H.}~\bibnamefont {Paukkunen}},\ and\
  \bibinfo {author} {\bibfnamefont {C.~A.}\ \bibnamefont {Salgado}},\ }\href
  {https://doi.org/10.1140/epjc/s10052-017-4725-9} {\bibfield  {journal}
  {\bibinfo  {journal} {The European Physical Journal C}\ }\textbf {\bibinfo
  {volume} {77}},\ \bibinfo {pages} {163} (\bibinfo {year} {2017})}\BibitemShut
  {NoStop}%
\bibitem [{\citenamefont {Dulat}\ \emph {et~al.}(2016)\citenamefont {Dulat},
  \citenamefont {Hou}, \citenamefont {Gao}, \citenamefont {Guzzi},
  \citenamefont {Huston}, \citenamefont {Nadolsky}, \citenamefont {Pumplin},
  \citenamefont {Schmidt}, \citenamefont {Stump},\ and\ \citenamefont
  {Yuan}}]{dulat2016new}%
  \BibitemOpen
  \bibfield  {author} {\bibinfo {author} {\bibfnamefont {S.}~\bibnamefont
  {Dulat}}, \bibinfo {author} {\bibfnamefont {T.-J.}\ \bibnamefont {Hou}},
  \bibinfo {author} {\bibfnamefont {J.}~\bibnamefont {Gao}}, \bibinfo {author}
  {\bibfnamefont {M.}~\bibnamefont {Guzzi}}, \bibinfo {author} {\bibfnamefont
  {J.}~\bibnamefont {Huston}}, \bibinfo {author} {\bibfnamefont
  {P.}~\bibnamefont {Nadolsky}}, \bibinfo {author} {\bibfnamefont
  {J.}~\bibnamefont {Pumplin}}, \bibinfo {author} {\bibfnamefont
  {C.}~\bibnamefont {Schmidt}}, \bibinfo {author} {\bibfnamefont
  {D.}~\bibnamefont {Stump}},\ and\ \bibinfo {author} {\bibfnamefont {C.-P.}\
  \bibnamefont {Yuan}},\ }\href
  {https://doi.org/http://dx.doi.org/10.1103/PhysRevD.93.033006} {\bibfield
  {journal} {\bibinfo  {journal} {Physical Review D}\ }\textbf {\bibinfo
  {volume} {93}},\ \bibinfo {pages} {033006} (\bibinfo {year}
  {2016})}\BibitemShut {NoStop}%
\bibitem [{\citenamefont {Pumplin}\ \emph {et~al.}(2002)\citenamefont
  {Pumplin}, \citenamefont {Stump}, \citenamefont {Huston}, \citenamefont
  {Lai}, \citenamefont {Nadolsky},\ and\ \citenamefont
  {Tung}}]{pumplin2002new}%
  \BibitemOpen
  \bibfield  {author} {\bibinfo {author} {\bibfnamefont {J.}~\bibnamefont
  {Pumplin}}, \bibinfo {author} {\bibfnamefont {D.~R.}\ \bibnamefont {Stump}},
  \bibinfo {author} {\bibfnamefont {J.}~\bibnamefont {Huston}}, \bibinfo
  {author} {\bibfnamefont {H.-L.}\ \bibnamefont {Lai}}, \bibinfo {author}
  {\bibfnamefont {P.}~\bibnamefont {Nadolsky}},\ and\ \bibinfo {author}
  {\bibfnamefont {W.-K.}\ \bibnamefont {Tung}},\ }\href
  {https://iopscience.iop.org/article/10.1088/1126-6708/2002/07/012/meta}
  {\bibfield  {journal} {\bibinfo  {journal} {Journal of High Energy Physics}\
  }\textbf {\bibinfo {volume} {2002}},\ \bibinfo {pages} {012} (\bibinfo {year}
  {2002})}\BibitemShut {NoStop}%
\bibitem [{\citenamefont {Armesto}(2006)}]{armesto2006nuclear}%
  \BibitemOpen
  \bibfield  {author} {\bibinfo {author} {\bibfnamefont {N.}~\bibnamefont
  {Armesto}},\ }\href {https://doi.org/10.1088/0954-3899/32/11/R01} {\bibfield
  {journal} {\bibinfo  {journal} {Journal of Physics G: Nuclear and Particle
  Physics}\ }\textbf {\bibinfo {volume} {32}},\ \bibinfo {pages} {R367}
  (\bibinfo {year} {2006})}\BibitemShut {NoStop}%
\bibitem [{\citenamefont {Thews}\ and\ \citenamefont
  {Mangano}(2006)}]{thews2006momentum}%
  \BibitemOpen
  \bibfield  {author} {\bibinfo {author} {\bibfnamefont {R.}~\bibnamefont
  {Thews}}\ and\ \bibinfo {author} {\bibfnamefont {M.~L.}\ \bibnamefont
  {Mangano}},\ }\href {https://doi.org/doi.org/10.1103/PhysRevC.73.014904}
  {\bibfield  {journal} {\bibinfo  {journal} {Physical Review C}\ }\textbf
  {\bibinfo {volume} {73}},\ \bibinfo {pages} {014904} (\bibinfo {year}
  {2006})}\BibitemShut {NoStop}%
\bibitem [{\citenamefont {Thews}\ \emph {et~al.}(2001)\citenamefont {Thews},
  \citenamefont {Schroedter},\ and\ \citenamefont {Rafelski}}]{thews}%
  \BibitemOpen
  \bibfield  {author} {\bibinfo {author} {\bibfnamefont {R.~L.}\ \bibnamefont
  {Thews}}, \bibinfo {author} {\bibfnamefont {M.}~\bibnamefont {Schroedter}},\
  and\ \bibinfo {author} {\bibfnamefont {J.}~\bibnamefont {Rafelski}},\ }\href
  {https://doi.org/10.1103/PhysRevC.63.054905} {\bibfield  {journal} {\bibinfo
  {journal} {Phys. Rev. C}\ }\textbf {\bibinfo {volume} {63}},\ \bibinfo
  {pages} {054905} (\bibinfo {year} {2001})}\BibitemShut {NoStop}%
\bibitem [{\citenamefont {Miller}\ \emph {et~al.}(2007)\citenamefont {Miller},
  \citenamefont {Reygers}, \citenamefont {Sanders},\ and\ \citenamefont
  {Steinberg}}]{miller2007glauber}%
  \BibitemOpen
  \bibfield  {author} {\bibinfo {author} {\bibfnamefont {M.~L.}\ \bibnamefont
  {Miller}}, \bibinfo {author} {\bibfnamefont {K.}~\bibnamefont {Reygers}},
  \bibinfo {author} {\bibfnamefont {S.~J.}\ \bibnamefont {Sanders}},\ and\
  \bibinfo {author} {\bibfnamefont {P.}~\bibnamefont {Steinberg}},\ }\href
  {https://doi.org/10.1146/annurev.nucl.57.090506.123020} {\bibfield  {journal}
  {\bibinfo  {journal} {Annu. Rev. Nucl. Part. Sci.}\ }\textbf {\bibinfo
  {volume} {57}},\ \bibinfo {pages} {205} (\bibinfo {year} {2007})}\BibitemShut
  {NoStop}%
\bibitem [{\citenamefont {Khachatryan}\ \emph {et~al.}(2017)\citenamefont
  {Khachatryan}, \citenamefont {Sirunyan}, \citenamefont {Tumasyan},
  \citenamefont {Adam}, \citenamefont {Asilar}, \citenamefont {Bergauer},
  \citenamefont {Brandstetter}, \citenamefont {Brondolin}, \citenamefont
  {Dragicevic}, \citenamefont {Er{\"o}} \emph
  {et~al.}}]{khachatryan2017suppression}%
  \BibitemOpen
  \bibfield  {author} {\bibinfo {author} {\bibfnamefont {V.}~\bibnamefont
  {Khachatryan}}, \bibinfo {author} {\bibfnamefont {A.}~\bibnamefont
  {Sirunyan}}, \bibinfo {author} {\bibfnamefont {A.}~\bibnamefont {Tumasyan}},
  \bibinfo {author} {\bibfnamefont {W.}~\bibnamefont {Adam}}, \bibinfo {author}
  {\bibfnamefont {E.}~\bibnamefont {Asilar}}, \bibinfo {author} {\bibfnamefont
  {T.}~\bibnamefont {Bergauer}}, \bibinfo {author} {\bibfnamefont
  {J.}~\bibnamefont {Brandstetter}}, \bibinfo {author} {\bibfnamefont
  {E.}~\bibnamefont {Brondolin}}, \bibinfo {author} {\bibfnamefont
  {M.}~\bibnamefont {Dragicevic}}, \bibinfo {author} {\bibfnamefont
  {J.}~\bibnamefont {Er{\"o}}}, \emph {et~al.},\ }\href
  {https://doi.org/http://dx.doi.org/10.1016/j.physletb.2017.04.031} {\bibfield
   {journal} {\bibinfo  {journal} {Physics Letters B}\ }\textbf {\bibinfo
  {volume} {770}},\ \bibinfo {pages} {357} (\bibinfo {year}
  {2017})}\BibitemShut {NoStop}%
\bibitem [{\citenamefont {ALICE-collaboration}(2014)}]{alice2014suppression}%
  \BibitemOpen
  \bibfield  {author} {\bibinfo {author} {\bibnamefont {ALICE-collaboration}},\
  }\href {https://doi.org/http://dx.doi.org/10.1016/j.physletb.2014.10.001}
  {\bibfield  {journal} {\bibinfo  {journal} {Physics Letters. B}\ }\textbf
  {\bibinfo {volume} {738}},\ \bibinfo {pages} {361–372} (\bibinfo {year}
  {2014})}\BibitemShut {NoStop}%
\bibitem [{\citenamefont {Sirunyan}\ \emph {et~al.}(2019)\citenamefont
  {Sirunyan}, \citenamefont {Tumasyan}, \citenamefont {Adam}, \citenamefont
  {Ambrogi}, \citenamefont {Asilar}, \citenamefont {Bergauer}, \citenamefont
  {Brandstetter}, \citenamefont {Brondolin}, \citenamefont {Dragicevic},
  \citenamefont {Er{\"o}} \emph {et~al.}}]{sirunyan2019measurement}%
  \BibitemOpen
  \bibfield  {author} {\bibinfo {author} {\bibfnamefont {A.~M.}\ \bibnamefont
  {Sirunyan}}, \bibinfo {author} {\bibfnamefont {A.}~\bibnamefont {Tumasyan}},
  \bibinfo {author} {\bibfnamefont {W.}~\bibnamefont {Adam}}, \bibinfo {author}
  {\bibfnamefont {F.}~\bibnamefont {Ambrogi}}, \bibinfo {author} {\bibfnamefont
  {E.}~\bibnamefont {Asilar}}, \bibinfo {author} {\bibfnamefont
  {T.}~\bibnamefont {Bergauer}}, \bibinfo {author} {\bibfnamefont
  {J.}~\bibnamefont {Brandstetter}}, \bibinfo {author} {\bibfnamefont
  {E.}~\bibnamefont {Brondolin}}, \bibinfo {author} {\bibfnamefont
  {M.}~\bibnamefont {Dragicevic}}, \bibinfo {author} {\bibfnamefont
  {J.}~\bibnamefont {Er{\"o}}}, \emph {et~al.},\ }\href
  {https://doi.org/https://doi.org/10.1016/j.physletb.2019.01.006} {\bibfield
  {journal} {\bibinfo  {journal} {Physics Letters B}\ }\textbf {\bibinfo
  {volume} {790}},\ \bibinfo {pages} {270} (\bibinfo {year}
  {2019})}\BibitemShut {NoStop}%
\bibitem [{\citenamefont {Acharya}\ \emph {et~al.}(2019)\citenamefont
  {Acharya}, \citenamefont {Acosta}, \citenamefont {Adamov{\'a}}, \citenamefont
  {Adolfsson}, \citenamefont {Aggarwal}, \citenamefont {Rinella}, \citenamefont
  {Agnello}, \citenamefont {Agrawal}, \citenamefont {Ahammed}, \citenamefont
  {Ahn} \emph {et~al.}}]{acharya2019Upsilon}%
  \BibitemOpen
  \bibfield  {author} {\bibinfo {author} {\bibfnamefont {S.}~\bibnamefont
  {Acharya}}, \bibinfo {author} {\bibfnamefont {F.~T.}\ \bibnamefont {Acosta}},
  \bibinfo {author} {\bibfnamefont {D.}~\bibnamefont {Adamov{\'a}}}, \bibinfo
  {author} {\bibfnamefont {J.}~\bibnamefont {Adolfsson}}, \bibinfo {author}
  {\bibfnamefont {M.~M.}\ \bibnamefont {Aggarwal}}, \bibinfo {author}
  {\bibfnamefont {G.~A.}\ \bibnamefont {Rinella}}, \bibinfo {author}
  {\bibfnamefont {M.}~\bibnamefont {Agnello}}, \bibinfo {author} {\bibfnamefont
  {N.}~\bibnamefont {Agrawal}}, \bibinfo {author} {\bibfnamefont
  {Z.}~\bibnamefont {Ahammed}}, \bibinfo {author} {\bibfnamefont {S.~U.}\
  \bibnamefont {Ahn}}, \emph {et~al.},\ }\href
  {https://doi.org/https://doi.org/10.1016/j.physletb.2018.11.067} {\bibfield
  {journal} {\bibinfo  {journal} {Physics Letters B}\ }\textbf {\bibinfo
  {volume} {790}},\ \bibinfo {pages} {89} (\bibinfo {year} {2019})}\BibitemShut
  {NoStop}%
\end{thebibliography}%
\end{document}